\documentclass[epsf,prb,twocolumn,showpacs,nofootinbib]{revtex4-1}

\usepackage[pdftex]{graphicx}
\usepackage{dcolumn}
\usepackage{bm}
\usepackage{epsfig}
\usepackage{latexsym}
\usepackage{amsmath}
\usepackage{amsfonts}
\usepackage{amssymb}
\usepackage{color}
\usepackage{array}
\usepackage{framed}

\begin{document}

\title{Entanglement entropy of $U(1)$ quantum spin liquids}
\author{Michael Pretko and T. Senthil\\
\emph{Department of Physics, Massachusetts Institute of Technology,
Cambridge, MA 02139, USA}}
\date{October 13, 2015}

\begin{abstract}

We here investigate the entanglement structure of the ground state of a (3+1)-dimensional $U(1)$ quantum spin liquid, which is described by the deconfined phase of a compact $U(1)$ gauge theory.  A gapless photon is the only low-energy excitation, with matter existing as deconfined but gapped excitations of the system.  It is found that, for a given bipartition of the system, the elements of the entanglement spectrum can be grouped according to the electric flux between the two regions, leading to a useful interpretation of the entanglement spectrum in terms of electric charges living on the boundary.  The entanglement spectrum is also given additional structure due to the presence of the gapless photon.  Making use of the Bisognano-Wichmann theorem and a local thermal approximation, these two contributions to the entanglement (particle and photon) are recast in terms of boundary and bulk contributions, respectively.  Both pieces of the entanglement structure give rise to universal subleading terms (relative to the area law) in the entanglement entropy, which are logarithmic in the system size ($\log L$), as opposed to the subleading constant term in gapped topologically ordered systems.  The photon subleading logarithm arises from the low-energy conformal field theory and is essentially local in character.  The particle subleading logarithm arises due to the constraint of closed electric loops in the wavefunction and is shown to be the natural generalization of topological entanglement entropy to the $U(1)$ spin liquid.  This contribution to the entanglement entropy can be isolated by means of the Grover-Turner-Vishwanath construction (which generalizes the Kitaev-Preskill scheme to three dimensions).

\end{abstract}
\maketitle

\normalsize

\section{Introduction}

In recent years it has become clear that some universal aspects of the ground states of interacting many particle systems can be fruitfully understood in terms of quantum entanglement in the corresponding wave function. 
For instance under a spatial bipartition 
fractional quantum Hall states have a universal  negative  constant  term  in their entanglement entropy (subleading  to the ubiquitous area law term).  This universal piece - known as the topological entanglement entropy - also appears in other phases of matter that have topological order  \cite{kitaev,levin}.  Examples are gapped quantum spin liquid phases of interacting quantum spins on a lattice. 
 Quantum spin liquids are fascinating states of matter which exhibit some of the most exotic phenomena of modern condensed matter physics.  They occur when there are large quantum fluctuations  of the spins which could possibly prevent ordering into a symmetry broken magnetic state even at zero temperature.  These large fluctuations prevent us from  usefully describing the system in terms  of semiclassical fluctuations of the original spin variables.  
 
 Quantum spin liquids come in many different varieties. In a  {\em gapped} quantum spin liquid, the bulk excitation spectrum has an energy gap. 
The low energy effective theory of  a wide class of such gapped spin liquid phases are  discrete gauge theories in their deconfined phase.  In this case the topological entanglement entropy may be viewed as a partial but universal characterization of such deconfined discrete gauge theories. 

In this paper we are concerned with the entanglement properties of a class of {\em gapless} quantum spin liquids. In contrast to their gapped cousins, theoretical understanding of  gapless quantum spin liquids is much less developed. Our focus in this paper is on a rather simple gapless quantum spin liquid state 
whose low energy effective theory is described by a deconfined $U(1)$ gauge theory.  These states of matter are known as $U(1)$ quantum spin liquids.  The excitation spectrum consists of one  gapless quasiparticle - identified with the photon of the gauge theory - and  other quasiparticle excitations that may be identified with electric and magnetic charges of the gauge theory.   We will here focus on spin liquids where the electric and magnetic charges are gapped, leaving the photon as the only gapless excitation.

In 2+1 dimensions, such a state with a gapless photon and gapped matter is unstable to confinement, as first demonstrated by Polyakov \cite{polyakov}.  However, in three or higher spatial dimensions, the $U(1)$ spin liquid can exist as a stable phase of matter. We shall  focus on the case of 3+1 dimensions, since that is the case of physical interest, though we shall also be able to make some statements about general dimensions.

These $U(1)$ quantum spin liquids are amongst  the simplest examples of gapless spin liquids, and many of their physical properties are well understood.  Amazingly the gaplessness of the photon is completely protected against small perturbations to the microscopic Hamiltonian including ones that break any global symmetry. This protection is a consequence of the non-local quantum correlations built into the ground state wave function of this spin liquid. These correlations enable the emergence of the deconfined $U(1)$ gauge theory and consequently the gapless photon. It is natural then to study the entanglement structure of the ground state wave function to characterize these quantum correlations. Our goal is to show how this entanglement structure is manifested in the entanglement entropy.  For such a gapless spin liquid is there an analogue  of the topological entanglement entropy that characterizes a gapped spin liquid?

Before stating our main result we briefly recall details of how the topological entanglement entropy appears in a topologically ordered state. Consider the entanglement entropy between two macroscopic regions of the system \cite{kitaev,levin}.  The leading behavior of the entanglement entropy is proportional to the area of the boundary between the regions.  The topological order then manifests itself (in two or three spatial dimensions) in the form of topological entanglement entropy, a negative subleading constant term:
\begin{equation}
S = \alpha L^{d-1} - \gamma + \cdot\cdot\cdot
\label{kit}
\end{equation}
for some nonuniversal constant $\alpha$ \cite{grover}.  (In general, the ``$\cdot\cdot\cdot$" can contain other subleading constant terms, not necessarily smaller than $\gamma$, particularly in three dimensions.  The topological $\gamma$ needs to be isolated by a special construction, to be described in Section 5.3, so care must be exercised in using Equation \ref{kit}).  As we shall discuss later, this subleading constant arises since the theory is one of closed loops.  The information that loops do not end in a region leads directly to a decrease in entropy.

We show in this paper that indeed the $U(1)$ spin liquid has an analogous universal signature of long-range entanglement in its subleading entanglement entropy behavior.  Unlike the subleading constant present in gapped phases, the subleading behavior in the $(3+1)$-dimensional $U(1)$ spin liquid takes the form of a logarithm: 
\begin{equation}
S = \alpha L^2 - (\gamma_{top} + \gamma_{ph}) \log L
\end{equation}
 The leading area law term is non-universal while the coefficient of the logarithm  is universal.  We will actually identify two universal subleading logarithms in the entanglement entropy which we have separated out as the two contributions $\gamma_{top}$ and $\gamma_{ph}$. The $\gamma_{ph}$ term originates from the gapless photon excitation and though universal is essentially local in character. It can be separated from the other contribution via a construction to be described in section 5.3.  The other subleading logarithm (the $\gamma_{top}$ term)  on the other hand, survives this procedure and is the natural generalization of topological entanglement entropy to this gapless phase.  For a connected entangling surface this ``topological" piece of the entanglement entropy, signifying the long-range entanglement of the system, simply has\cite{foot}:
\begin{equation}
\gamma_{top} = 1
\end{equation}
 On the other hand, $\gamma_{ph}$ depends on the shape of the entangling surface.  But if we consider as an example a simple spherical entangling surface, we have $\gamma_{ph} = \frac{1}{45}$  so that the net coefficient of the logarithm is $\frac{46}{45}$.

The low energy deconfined $U(1)$ gauge theory arises as an emergent property\cite{forster,wen} of one possible phase of a microscopic spin (or boson) system. In particular the microscopic Hilbert space is simply that of a tensor product of spin states (or boson states) at various lattice sites.   It is important to emphasize that concrete microscopic model Hamiltonians in 
a variety of lattice spin systems (or closely related models of interacting bosons) have been demonstrated to be in such liquid phases where a deconfined $U(1)$ gauge theory emerges\cite{levinu1f,bosfrc3d,hfb04,3ddmr,lesikts05,kdybk}.  Future numerical studies of these models could possibly detect the universal structure in the entanglement entropy we find.

Besides being of theoretical interest as a natural setting for studying gapless long-range entanglement, the $U(1)$ spin liquid is also a candidate to describe physical spin liquids in
``quantum spin ice" materials\cite{balentsqspice}. 
And for high-energy physicists, we remark that, depending on one's biases about the fundamental Hilbert space leading to the Standard Model, the calculation in this paper could be applicable to understanding the entanglement structure of our universe.  After all, if we ignore gravity for the moment, the low energy theory of our universe is a deconfined $U(1)$ gauge theory, and all known matter fields have a nonzero mass.  This requires some acceptance of a picture of emergence from a local tensor product Hilbert space, but the possibility seems worth considering.

Previously\cite{xy,qi} the entanglement entropy of gapless quantum spin liquid systems described as deconfined discrete gauge theories with gapless matter fields in two dimensions was computed. A separation between a gapless contribution and topological contribution to the entanglement entropy  was argued on general grounds. Our result in the present paper bears some resemblance though the details are different.

The present work will also provide a more physical perspective on some of the issues plaguing the concept of entanglement in gauge theories.  The central problem is that the set of gauge-invariant states does not possess a tensor product Hilbert space structure, which seems to be a prerequisite for a sensible notion of entanglement.  Thus, even providing a $definition$ of entanglement (let alone its calculation) has been a source of headaches for the high-energy community.  One common line of thought is that we must resort to complex algebraic procedures to define entanglement\cite{casini}.  More recently, References \onlinecite{ghosh} and \onlinecite{aoki} have proposed adding in non-gauge-invariant degrees of freedom as a minimal way to embed the system in a tensor product Hilbert space.  While such a procedure may at first seem a little ad hoc in a high-energy context, the condensed matter perspective on the problem makes it quite natural.  As we shall discuss below, these ``non-gauge-invariant" degrees of freedom simply represent particles coupled to the gauge field.  For a gauge theory emerging from a local boson system, such as would occur in a solid, gauge fields and particles always emerge jointly in this fashion, so it would be unphysical to consider entanglement in the gauge theory without taking the particles into account.  Thus, from our condensed matter perspective, these additional degrees of freedom are inevitable, and the issues plaguing the high energy community are not a concern.

As this draft was being prepared, we became aware of recent work \cite{radi} in the high-energy community which reaches many of the same conclusions, though through a different methodology.  The fact that these two very different perspectives on the problem yield similar results for the entanglement structure is encouraging.

\section{Definition of the Problem}

We consider a system whose low-energy spectrum is described by the deconfined phase of a compact $U(1)$ gauge theory, so that its only low-energy excitation is the gapless photon, with all matter fields gapped.  We will mainly work on a spatial lattice, since the study of entanglement requires proper short-distance regularization, but none of the general conclusions will depend on a particular choice of lattice structure, and connection with the continuum theory will be made.  Since entanglement is most naturally defined on constant time slices of spacetime, we will not need to make any use of the time coordinate in the analysis.

A compact $U(1)$ gauge theory on the lattice can be described by a $U(1)$ variable $e^{ia_{ij}}$ (essentially a quantum rotor) living on each link of the lattice, where $i$ and $j$ denote the two endpoints of the link.  Since the angular variable $a_{ij}$ representing the spatial components of the vector potential is compact ($i.e.$ only values between 0 and $2\pi$ are distinct), the conjugate (momentum) variable corresponding to $a_{ij}$ has integer eigenvalues, as is usual for an angular momentum variable.  For the case of the vector potential in the $U(1)$ gauge theory, the conjugate variable corresponds to the electric field $E$ living on each link.  Thus, for a compact $U(1)$ gauge theory, the electric field on each link is quantized to integer multiples of some specific value, taken to be 1 for convenience.  This quantum rotor language is essentially equivalent to working with large $S$ spins.  However, this is just a high-energy detail, and even a simple spin-$1/2$ system can flow towards this rotor description under the renormalization group.\cite{hfb04}

At the field theory level, one usually also has a timelike component $a_0$ of the vector potential.  However, this variable does not represent an independent degree of freedom, but rather a sort of ``Lagrange multiplier" variable enforcing the Gauss's law constraint ($\nabla\cdot E = 0$) on our low-energy Hilbert space, and a proper treatment of entanglement should start from the Hamiltonian formulation of lattice gauge theory, where $a_0$ is integrated out in favor of the Gauss's law constraint.  The procedure is standard and is reviewed in Appendix A.  Furthermore, this perspective on the Hilbert space is very natural for the case of gauge fields emerging from bosonic models, as happens with spin liquids.  After integrating out $a_0$, the continuum low-energy effective Hamiltonian for the $U(1)$ spin liquid phase becomes:
\begin{equation}
H = \int d^3x\bigg[\frac{1}{2}(E^2+B^2) + U(\nabla\cdot E)^2\bigg]
\end{equation}
where the first two terms represent the standard Hamiltonian for electromagnetism, with $B$ representing the curl of $a$.  (On the lattice, $E$ naturally lives on the links and $B$ will naturally live on the plaquettes.)  The last term serves to enforce the gauge constraint on the low-energy subspace, effectively serving as an energy penalty for non-gauge-invariant states.  This last term is usually not written explicitly in a Maxwell theory, but we retain it here, since we will see shortly that it represents the energy gap to particle states.

Thus, we take our underlying Hilbert space to be that of quantum rotors $e^{ia_{ij}}$ living on all the spatial links of our lattice.  We can label the states of a Hilbert space for each link by the set of integers, corresponding to the quantized electric field values, $|E=n\rangle$.  We then take the Hilbert space of the whole system to be the tensor product of the Hilbert spaces for the individual links.  As a convenient pictorial representation, we can equivalently think of this as the Hilbert space of directed strings on the lattice, representing electric field lines.  For each link, we regard the state $|0\rangle$ as the absence of a string on that link.  We can visualize the state $|1\rangle$ as a directed string running along the link in a specified direction.  (This direction should be specified consistently for all links of the lattice with the same orientation.)  The state $|-1\rangle$ then corresponds to a directed string in the opposite direction.  All of the other states $|\pm n\rangle$ for $n>1$ can be regarded as $n$ strings overlapping on the same link, all pointing in the same direction.  The ground state of our theory, which has all matter fields gapped, will be made up of states which satisfy the source-free Gauss's law, $\nabla\cdot E = 0$, meaning that no net flux flows into or out of a specific site of the lattice, so the electric field lines must form closed loops (bearing in mind that loops may overlap).  The total Hilbert space of our system corresponds to that of open and closed strings on the lattice, while the ground state occupies the sector of Hilbert space made up only of closed-loop configurations.

\begin{figure}[b!]
 \centering
 \includegraphics[scale=0.4]{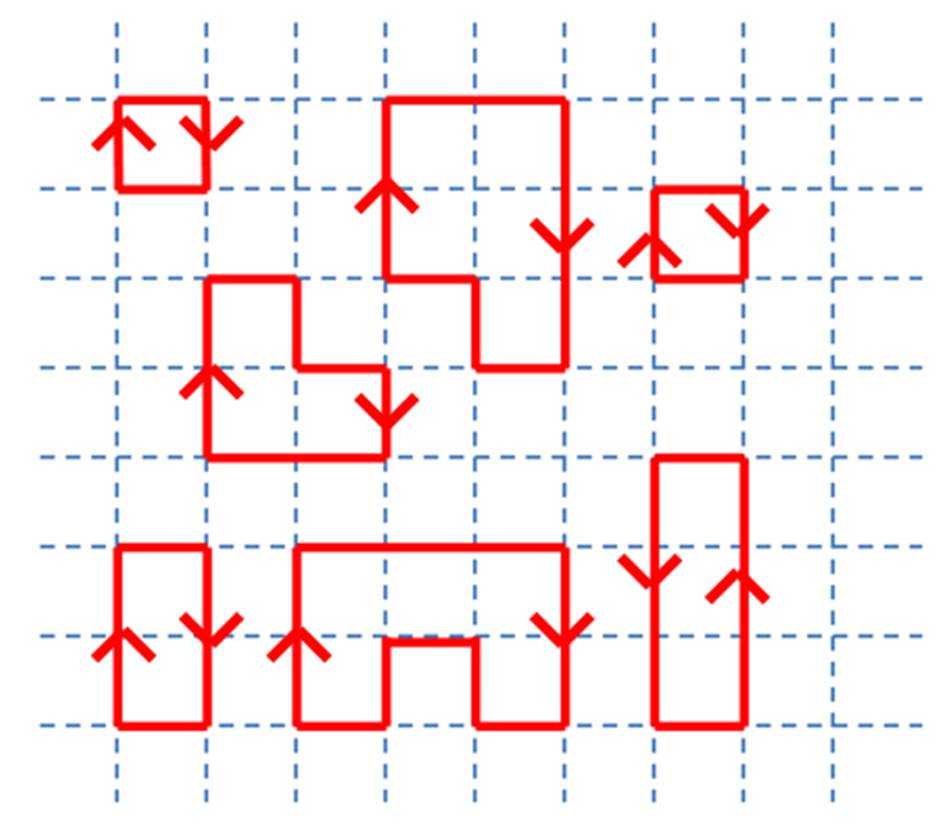}
 \caption{A typical closed-loop configuration.  The ground state of the $U(1)$ spin liquid will be a superposition of such configurations.}
 \label{fig:compare}
 \end{figure}

\begin{figure}[t!]
 \centering
 \includegraphics[scale=0.4]{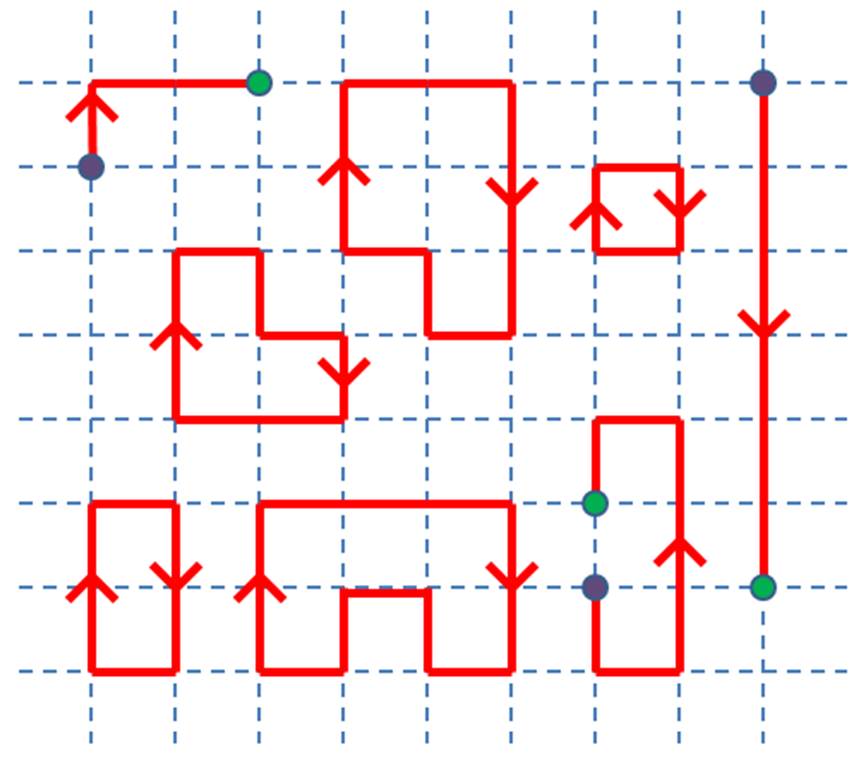}
 \caption{A typical configuration with open strings.  The endpoints of the strings represent matter fields.  States composed of open strings are taken to be gapped in the $U(1)$ spin liquid phase.}
 \label{fig:compare2}
 \end{figure}

Some may regard this Hilbert space as too big.  The gauge-invariant states, including the ground state of the $U(1)$ spin liquid, correspond to closed loops of strings.  Indeed, there is no way to endow the Hilbert space of only closed loops with a local tensor product structure.  This leads to the conventional wisdom that gauge theories do not have a natural tensor product structure, and one must resort to more detailed mathematical procedures to define entanglement \cite{casini}.  However, the tensor product issue is easily solved by simply working in the Hilbert space of both open and closed strings, which does have a local tensor product structure, as described above and in previous works\cite{ghosh,aoki}.  The ``non-gauge-invariant" states, corresponding to open strings, can simply be regarded as states with matter present at the endpoints, \emph{defining} matter as the endpoints of strings, as is common in string-net models \cite{xiao}, and more generally in the context of emergent gauge theories in spin liquids and other bosonic models.  In field theoretic contexts, one usually introduces a separate matter field coupled to the gauge field, and then enforces gauge invariance by requiring that the strings only end on particles of the matter field.  If we did this, we would then have a matter field Hilbert space, a gauge-field Hilbert space, and an unsightly constraint between the two which prevents the system from having a tensor product structure.  However, we can eliminate the need for such a constraint and restore the local tensor product nature by interpreting open strings as pairs of particles.  The redundancy implied by gauge invariance is then simply the redundancy of the matter field itself.  For field theoretic calculations, representing matter by independent fields is quite useful, but in principle it is unnecessary to have a description of matter independent of the strings.  For the purposes of investigating entanglement, it is simpler to do away with separate matter fields, regarding a gauge theory simply as a theory of both open and closed strings.\cite{foot2}  This perspective on gauge theory should seem natural to those familiar with lattice models for discrete gauge theories, such as the toric code (a $\mathbb{Z}_2$ gauge theory).  It has even been shown that the Hilbert space of non-abelian gauge theories can be thought of in the same fashion \cite{aspect} (though of course these theories are much more susceptible to confinement).
 
It should be noted that one will run into situations of trying to define entanglement for gauge theories where the number of matter species is greater than the number of endpoints of strings.  This corresponds to the matter particles having some extra internal structure, such as spin, or having different ``flavors" (such as electron, muon, etc., in the Standard Model).  Such internal structure to matter requires either additional structure to the theory beyond the $U(1)$ framework, such as the electroweak structure of the Standard Model, or some dynamically generated extra structure to the theory.  For example, the dynamics of the electric strings could be such that they tend to form binary bound states (``ribbons"), providing an extra orientational degree of freedom in the low-energy Hilbert space \cite{ribbon}.  However, such extra structure should not lead to any significant alteration of the conclusions found here, as we shall see that the important piece of the entanglement entropy is dictated purely by the closed loop constraint of the ground state, a fact which is not changed when the loops carry extra structure.

The above prescription gives us a way to describe the local tensor product structure of the Hilbert space of the compact $U(1)$ gauge theory in any phase, whether the matter fields (string endpoints) are gapless, gapped, or even confined.  We will now, however, focus our attention on a $U(1)$ spin liquid which has its matter fields deconfined, but gapped out to high energies.  In other words, large loops have proliferated, but there is a large energy penalty for open strings, so the ground state is simply a superposition of configurations of closed loops (not necessarily with equal weight).  The restriction to closed loops usually has interesting manifestations in the entanglement structure of a theory.  For example, the deconfined phase of a $Z_2$ gauge theory, which has a ground state in which closed loops have proliferated, has a topological entanglement entropy of $-\log 2$, essentially arising from the restriction of the Hilbert space to closed loops \cite{kitaev,levin}.  We will here investigate the possibility of analogous entanglement effects in the deconfined phase of a $U(1)$ gauge theory, where closed loops have similarly proliferated.

However, the $U(1)$ case has an important difference from the $Z_2$ (or any discrete) gauge theory.  In a discrete gauge theory, there are a finite number of values that can be taken on each link.  In a $Z_n$ gauge theory for example, $n$ loops sitting on top of each other is equivalent to the trivial configuration.  This is closely related to the fact that charge is only conserved mod $n$ in such a theory.  In the $U(1)$ gauge theory, on the other hand, charge is conserved absolutely, and each link has an infinite-dimensional Hilbert space, corresponding to arbitrarily high values of the electric field, though large values of the electric field will generally be energetically unfavorable and therefore suppressed in the wavefunction.  This corresponds to a repulsion between loops running in the same direction.

In the end, we will find that, while a gapped topologically ordered phase is characterized by a universal subleading constant in the ground state entanglement entropy, the deconfined $U(1)$ phase is characterized by two separate universal subleading logarithmic terms in the ground state entanglement entropy.  One contribution will come from the low energy conformal field theory of the photon and will be essentially local in character.  The other will come from the closed loop constraint and can be associated with the particle structure of the theory.  It corresponds to the constraint of zero electric flux through any closed surface, in agreement with logic put forward in a previous construction of a gapless spin liquid state \cite{wen2}.  This ``particle" contribution will be seen to be the natural generalization of topological entanglement entropy to the $U(1)$ spin liquid.  There is a clean separation between the ``topological" piece, coming from the gapped particles, and the conformal piece, coming from the gapless photon.  Such separation between a gapless contribution and a topological contribution to the entanglement entropy is reminiscent of that observed before\cite{xy,qi}  for emergent discrete gauge theories coupled to gapless matter fields.  However, counter-examples are known which do not possess this separability property \cite{sachdev}.  The differences between these models are discussed further in Appendix E.

\section{Entanglement Spectrum Classified by Boundary Conditions}

We now partition our system into two macroscopic regions, and we investigate the entanglement between the two regions.  For a given wavefunction on a system with specified partition, finding the entanglement spectrum is equivalent to finding the Schmidt decomposition of the wavefunction between the two regions:
\begin{equation}
|\Psi\rangle = \sum_n e^{-\lambda_n/2} |\psi_n\rangle_A |\phi_n\rangle_B
\label{schmidt}
\end{equation}
where $\{\psi_n\}$ and $\{\phi_n\}$ are bases for systems $A$ and $B$ respectively, and the values $\lambda_n$ make up the entanglement spectrum.  While the closed loop constraint might at first seem like it would be an additional complication, it is actually a drastic simplification when it comes to finding the Schmidt decomposition.  While the state of the whole system is made up of closed loops, a state $|\psi_n\rangle_A$ defined on $A$ can have loops seemingly end on the boundary, so long as the partnered state $|\phi_n\rangle_B$ defined on $B$ picks up where $A$ leaves off and continues the strings running in the same direction.  In other words, for any closed loop wavefunction, if we know that a state on system $A$ has a specified flux configuration passing through the boundary, then the state on $B$ must have those same boundary conditions.  Furthermore, the states $|\psi_n\rangle_A$ and $|\phi_n\rangle_B$ cannot have a superposition of different boundary conditions without leading to a mismatch of boundary conditions between the two sides.

\begin{figure}[b!]
\centering
 \includegraphics[scale=0.45]{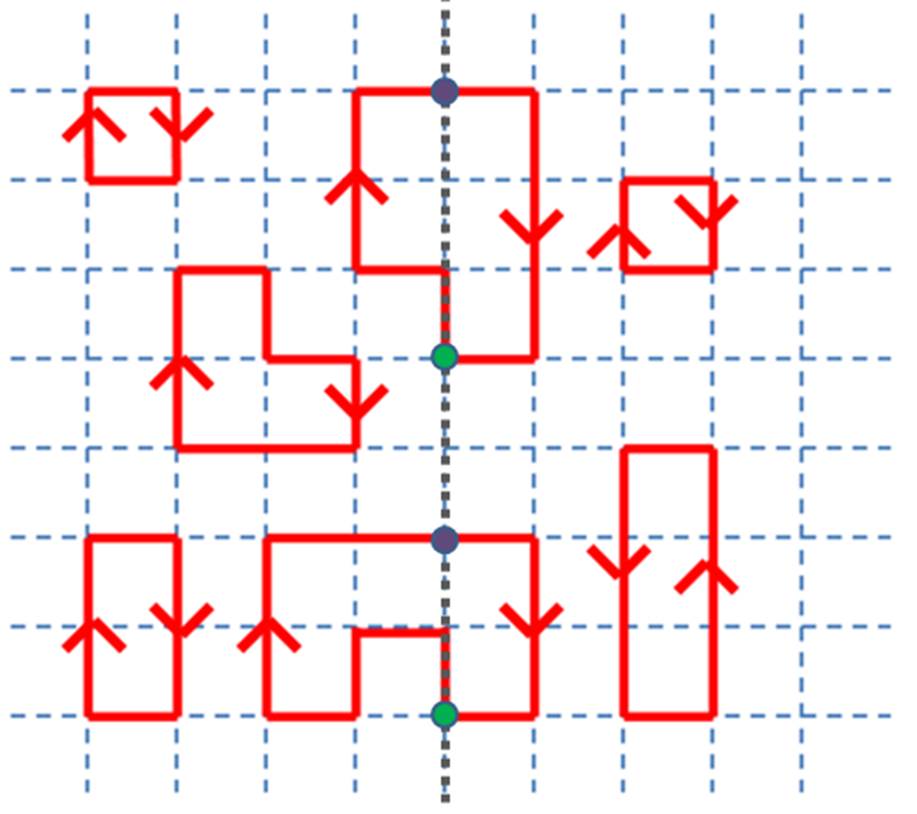}
 \caption{When we partition the system into two regions, each configuration has a corresponding configuration of flux points between the two regions.  The closed loop constraint ensures that the boundary conditions must match on the two sides.  We define a partition of the system as a partition of the links, and the links on the partitioning line above have been taken as part of the left subsystem to avoid ambiguity in defining the location of the flux points.}
 \label{fig:partition}
 \end{figure}

To phrase this more rigorously, consider a general wavefunction in this Hilbert space, which can be written as:
\begin{equation}
|\Psi\rangle = \sum_{nm} C_{nm}|\psi_n\rangle_A |\phi_m\rangle_B
\end{equation}
for some matrix of coefficients $C_{nm}$.  By performing a singular value decomposition on $C_{nm}$, one obtains the Schmidt decomposition in Equation \ref{schmidt}.  Now let the basis elements $\{|\psi_n\rangle\}$ and $\{|\phi_m\rangle\}$ be specified by their electric field eigenvalues on each link (which are restricted to be integers).  Thus, each basis element has a specified electric flux through the boundary.  Because we take our ground state wavefunction to be a superposition of closed loops, $C_{nm}$ is only nonzero when $|\psi_n\rangle_A$ and $|\phi_m\rangle_B$ have matching boundary conditions, in the sense that a unit of flux leaving $A$ corresponds to a unit of flux entering $B$ at the corresponding point on the boundary.  Thus, we see that $C_{nm}$ has a block diagonal form, with blocks corresponding to different electric boundary conditions between the two systems.  The Gauss's law constraint of closed loops ensures that there is no mixing between blocks.  Then, in order to obtain the Schmidt decomposition, we can perform a singular value decomposition separately on each of the boundary condition blocks.  The squared magnitude of the diagonal elements in this decomposition then immediately yield the eigenvalues of the reduced density matrix for one of the subsystems, $i.e.$ the entanglement spectrum.  Due to the block diagonal form, each element of the entanglement spectrum corresponds to a specific set of boundary conditions between the two regions, which will shortly allow us to match up part of the entanglement spectrum with a thermal spectrum of particles living on the boundary.

This decomposition of the entanglement spectrum into different flux sectors has previously been noticed by at least three other groups \cite{aoki,ghosh,don2}.  Furthermore, others have already found evidence of this ``boundary theory" in the entanglement spectrum, both through formal arguments \cite{don2,donnelly} and through analysis of specific wavefunctions \cite{buiv}.  Also, some work has been done on extracting universal logarithmic terms in the entanglement entropy \cite{chm,dowker,kuo}.  In the present work, we shall go two steps further.  First, we will use a simple thermodynamic picture to separate the entanglement entropy into a boundary particle contribution and a bulk photon contribution.  Second, we will make use of a special construction (see Section 5.3) to separate the universal pieces of the two contributions, identifying the particle contribution as the gapless analogue of topological entanglement entropy, while the photon contribution is essentially local in character.  Furthermore, the topological piece will be seen to be a direct consequence of neutrality in the boundary particle gas, which corresponds to the closed loop constraint on the ground state wavefunction.

\begin{figure}[t!]
\centering
 \includegraphics[scale=0.4]{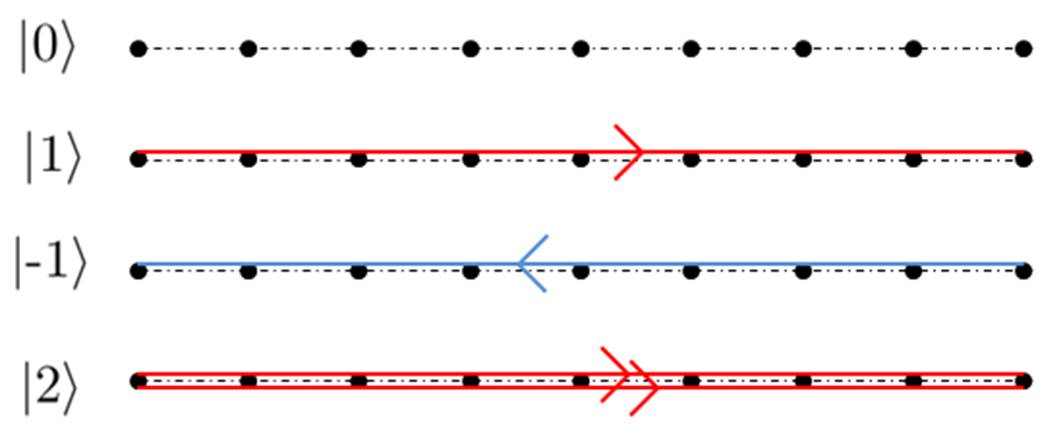}
 \caption{In a one-dimensional chain, the only states consistent with Gauss's law have uniform electric flux throughout the whole chain.  These states can be labeled by integers, representing the quantized value of the electric flux.  Since the flux is uniform throughout the whole system, in particular it represents the flux across the boundary between any two partitions of the system.}
 \label{fig:oned}
 \end{figure}

As a very simple example of the decomposition into boundary sectors, consider a one-dimensional chain of links, as illustrated in Figure \ref{fig:oned}.  Such a one-dimensional gauge theory will be trivially confining (the only gauge-invariant quantity which can appear in the Hamiltonian is the electric field), so we expect nothing interesting in the entanglement structure.  Nevertheless, the one-dimensional Hilbert space exhibits the boundary decomposition in a particularly simple way.  Each site touches precisely two links, so any flux carried into the site by one link must be carried out in the same amount by the other link.  In other words, the only gauge invariant states are those with uniform electric field across the whole chain, $|E=n\rangle = |n\rangle_A|n\rangle_B$, which is a direct product state.  A general gauge invariant wavefunction in one dimension can then be written as:
\begin{equation}
|\Psi\rangle_1 = \sum_n c_n |n\rangle_A |n\rangle_B
\end{equation}
which is already in Schmidt form, with the elements of the entanglement spectrum labeled by the integer $n$, which is precisely the electric flux through the boundary between the regions.  Thus, in one dimension, any wavefunction has entanglement spectrum in one-to-one correspondence with electric boundary conditions.  Of course, we expect that uniform electric fields throughout the system will be energetically costly, so the one-dimensional gauge theory ought to have $|E=0\rangle = |0\rangle_A |0\rangle_B$ as its ground state.  Thus, in one dimension, the entanglement entropy of a $U(1)$ gauge theory is zero, a fairly uninteresting result, in accordance with the statement that the one-dimensional gauge theory is confining.

However, our main interest will be deconfined phases, for which we must go to higher spatial dimensions.  In higher dimensions we must now ask, given a labeling of the entanglement spectrum by boundary conditions, have we completely labeled the entanglement spectrum, as in one dimension?  Each element of the entanglement spectrum corresponds to a specific boundary condition, but does each boundary condition define a single element in the entanglement spectrum?  In general, the answer is no.  However, we can gain intuition for the general situation by first examining the class of wavefunctions that do satisfy this criterion.  Consider first a direct product wavefunction, which has no correlation between the two sides, $|\Psi\rangle = |\psi\rangle_A |\phi\rangle_B$.  Such a wavefunction in general cannot satisfy the Gauss's law constraint, since there are no correlations between boundary conditions on the two sides.  Furthermore, such a wavefunction is not guaranteed to be free of magnetic monopoles.  In order to obtain a wavefunction consistent with our constraints, we must project into the sector of zero electric and magnetic charge.  As shown in Appendix B, the projection of such a direct product state onto the zero particle sector results in a state with entanglement spectrum in one-to-one correspondence with the electric boundary conditions.  The reduced density matrix for one of the regions will then simply describe a classical probability distribution for the electric flux on the boundary.  Since the electric flux is quantized, we can think of this as a classical theory of charged particles on the boundary, associating positive charge to one orientation of flux and negative charge to the other.  (One might at first think (as we did) that we could simultaneously specify both electric and magnetic boundary conditions between the two regions, ending up with a boundary theory of both electric and magnetic charges.  But in fact, the electric and magnetic perspectives are complementary, rather than additive, and it is sufficient to consider the electric boundary theory.  This point is discussed further in Appendix C.)

However, the form of the wavefunction chosen above, the projection of a direct product onto the zero charge sector, assumes that the only correlations in our system come from the local neutrality constraint.  Such a wavefunction is often appropriate for gapped topological phases.  For example, in a two-dimensional toric code model, which characterizes the deconfined phase of a $Z_2$ gauge theory, the wavefunction is given by an equal weight superposition of all closed loop configurations \cite{fradkin}.  This is equivalent to the projection of the superposition of all open and closed string configurations (a direct product state) into the closed loop sector.  Thus, the deconfined $Z_2$ gauge theory will be well described by a projected direct product state.  But while such states are appropriate for dealing with gapped topological order, in the case at hand of a deconfined $U(1)$ gauge theory, we know that there is also a gapless photon, which will lead to long-range correlations in the system independent of the local neutrality constraint.  We therefore expect that, in the presence of gapless modes, there is no reason that the entanglement spectrum should be in one-to-one correspondence with boundary conditions.

Despite the labeling no longer being one-to-one, the decomposition into boundary sectors still holds and is still quite useful.  One can still think of this as a theory of boundary particles, except that now each particle configuration is carrying extra internal entropy due to the gapless photon.  Let us denote the eigenvalues of our reduced density matrix as $p_{bc,i}$, where $bc$ labels the boundary condition sector and $i$ runs over all values within that boundary sector.  We further denote $P_{bc} = \sum_i p_{bc,i}$, which essentially represents the probability of a specific configuration of particles in the boundary theory.  We can further define $p'_{bc,i} = p_{bc,i}/P_{bc}$, which satisfies $\sum_i p'_{bc,i} = 1$.  Thus, $p'_{bc,i}$ represents the probability distribution of internal configurations, given a fixed boundary condition.  The entanglement entropy of the system can now be written as:
\begin{align}
\begin{split}
S = -\sum_{bc,i} p_{bc,i}\log p_{bc,i} = &-\sum_{bc,i}P_{bc}p'_{bc,i}\log(P_{bc}p'_{bc,i})\\
= -\sum_{bc}P_{bc}\log P_{bc} &-\sum_{bc}P_{bc}\sum_i p'_{bc,i}\log p'_{bc,i} \\
\equiv S_{bc} +& \sum_{bc}P_{bc}S_{int,bc}
\label{break}
\end{split}
\end{align}
The first term is simply the entropy of the distribution of boundary conditions, corresponding to the entropy of a theory of charged particles.  The second term represents the average internal entropy carried by each particle configuration.  Provided that the internal entropy does not depend strongly on the choice of boundary conditions ($i.e.$ $S_{int,bc}$ are all close to some mean value $S_{int}$ for most boundary conditions), we can simply write:
\begin{equation}
S = S_{bc}+S_{int}
\end{equation}
where the first term comes from the entropy of the particle distribution, and the second is the internal entropy.  The nature of this internal entropy, and the justification for $S_{int,bc}$ having small fluctuations about its mean, requires a different perspective on the problem, which we shall explore in the next section.

\section{The Bisognano-Wichmann Theorem}

In searching for the correct description of entanglement in the $U(1)$ spin liquid, we are greatly aided by the fact that the deconfined $U(1)$ gauge theory has an emergent relativistic symmetry, since its low-energy effective theory is described by standard noncompact electrodynamics in $d+1$ dimensions:
\begin{equation}
S = \frac{1}{2e^2}\int d^{d+1}x F^{\mu\nu}F_{\mu\nu}
\label{elec}
\end{equation}
where the field strength as usual is $F_{\mu\nu} = \partial_\mu a_\nu - \partial_\nu a_\mu$.  (In three spatial dimensions, our case of primary interest, one should also allow for the possibility of a $\theta$ term in the action, $\theta\epsilon^{\mu\nu\lambda\sigma} F_{\mu\nu}F_{\lambda\sigma}$.  This possibility is discussed further in Appendix D, where it will be seen that, except for minor subtleties, its presence will not alter the conclusions reached here by taking $\theta = 0$.)  For a theory with such relativistic invariance, the entanglement spectrum can be calculated exactly for the simple geometry of a planar partitioning surface.  Consider a system described by a local Hamiltonian density, such that the Hamiltonian is given by $H = \int d^dx \mathcal{H}$.  The Bisognano-Wichmann result \cite{wichmann,bianchi} states that in a relativistic system (with units chosen such that the speed of light is $1$), for a planar partitioning surface at $x_1=0$, the reduced density matrix describing $x_1>0$ is given by:
\begin{equation}
\rho \propto \exp\bigg(-\int_{x_1>0} d^dx (2\pi x_1)\mathcal{H}\bigg)
\end{equation}
The entanglement Hamiltonian density is given by the real Hamiltonian density, except with an extra position-dependent weighting factor.  How should we interpret this entanglement spectrum, and how can we extract important quantities like the entanglement entropy?  As has been discussed elsewhere \cite{wong,me}, this reduced density matrix essentially describes a local thermal equilibrium.  Note that the density matrix has the form $\rho \propto \exp(-\int dx \beta(x)\mathcal{H})$.  For constant $\beta$, this density matrix would exactly describe a thermal system at temperature $\beta^{-1}$.  When $\beta$ is nonuniform, we can still usefully think of this as a thermal distribution, but now with a locally defined temperature $T(x)=\frac{1}{2\pi x_1}$, regarding the system as being in local thermal equilibrium, in a somewhat similar manner to a Thomas-Fermi approximation.  The temperature cools off to zero far away from the edge, but it reaches arbitrarily high values close to the partition.  This makes some intuitive sense, in that it is degrees of freedom closest to the edge which are most affected by the tracing out procedure.

To demonstrate the essential correctness of this interpretation, note that it can be used to reproduce exactly a well-known formula for $1+1$ dimensional conformal field theories, as first noticed in Reference \onlinecite{wong}.  Consider such a CFT characterized by central charge $c$.  The thermal entropy density of such a system is given\cite{fradkin} by $s(T) = \frac{\pi c}{3}T$.  Now suppose our partition is into a finite segment of length $L$ and its exterior.  We can get the entanglement entropy by integrating the local thermal entropy density over the two edges of the system:
\begin{equation}
S = 2\int_{x>0} dx \frac{\pi c}{3} \frac{1}{2\pi x} = \frac{c}{3}\log(L/a)
\label{cft}
\end{equation}
where the factor of two comes from the the two edges of the system, and we have cut off the long-distance divergence at a distance of order $L$.  The short-distance cutoff $a$ represents the lattice scale.  This formula exactly reproduces the known result for entanglement entropy in a 1+1 dimensional CFT \cite{fradkin}.  For any other relativistic theory, the procedure is much the same.  We can simply integrate the local thermal entropy over the interior of the partitioning surface.  In dimensions higher than one, the factor of two in Equation \ref{cft} is generalized to a factor of the surface area of the partitioning surface, leading to a natural understanding of the area law (with area law violations stemming from divergences in the $x_1$ integral, as above).  Even for certain non-relativistic systems this procedure is useful.  For example, the Bisognano-Wichmann result and local thermal approximation can be used in the context of Fermi surface systems to yield the Widom formula for Fermi surface entanglement entropy \cite{me}.  In general, for any model, there should be corrections due to the fact that one's choice of partitioning surface is usually not strictly planar and also due to the temperature gradient, but the essential physical picture of local thermodynamics seems to be a valid one.

It should be noted that there is a subtlety in applying this procedure to a lattice system, which is our case of physical interest.  Our system is only truly relativistic at low energies.  At high energies, obviously the presence of a spatial lattice breaks the relativistic invariance.  Since entanglement metrics, such as entanglement entropy, are often sensitive to short-distance physics, it is questionable to apply the result directly to the lattice.  Furthermore, the local thermal picture is sensitive to excitations at arbitrarily high temperatures, so one should really have well-defined relativistic excitations at all scales in order to apply this procedure.  The trick is to apply the Bisognano-Wichmann theorem and the local thermal approximation not to the lattice system directly, but rather to a UV complete relativistic theory which reduces to Maxwell theory at low energies.  This is easily accompished by regarding the $U(1)$ gauge theory as descending from a non-abelian gauge theory via symmetry breaking.  This change of the UV behavior of our theory can affect non-universal physics, such as the coefficient of the area law.  However, the important point is that the universal physics is independent of the short-distance regularization (by definition).  In this work, we will only focus on universal quantities, which do not depend on short-distance physics, so this change of the UV will not be important.  More detailed discussion of the embedding into a UV complete theory can be found in Appendix C.  This procedure is very important conceptually, but for most practical purposes, we can simply proceed with the low-energy Maxwell theory.

In order to calculate the entanglement entropy for the $U(1)$ spin liquid via the local thermal method, we must consider what degrees of freedom we have which will contribute to the thermal entropy density.  Obviously we have the gapless photon, representing the fluctuations of closed strings.  By dimensional analysis, this contribution to the entropy density must scale as $T^3$, so its contribution to the entanglement entropy falls off as $x^{-3}$ as we move away from the partitioning surface.  This is a slow power-law decay, so thermal excitations of photons exist well into the bulk of the region.

However, there is an additional set of degrees of freedom which we must consider.  For a gauge theory ``without matter fields," as described by Equation \ref{elec}, the traditional Hilbert space considered is that of gauge-invariant (closed loop) states.  However, as we have argued at the beginning of the present work, a natural definition of entanglement leads us to also include the ``non-gauge-invariant" open string states, which can be simply interpreted as electric particles, which we take to be gapped.  By universality, the details of the gapped charged sector should not be important.  It is therefore convenient to incorporate these gapped electric charges into a relativistic quantum field theory   so that we can continue to use the BW theorem. To that end, we consider a theory with a relativistic Lagrangian 
\begin{equation}
\label{u1lag}
{\cal L} = {\cal L}[\psi, a_\mu] + \frac{1}{2e^2} F_{\mu \nu} F^{\mu \nu}
\end{equation}
Here $\psi$ is the massive field describing electric charges that couple minimally with the $U(1)$ gauge field $a_\mu$.  We may take $a_\mu$ to be non-compact as befits the low energy effective theory of the $U(1)$ spin liquid. However the emergence of this spin liquid from an underlying lattice spin system means that there will inevitably be magnetic monopoles in the spectrum.  The above non-compact theory should be regarded as an effective theory obtained by integrating out the monopoles.  Equivalently, since we are interested in universal aspects of the physics, we may consider a model where the monopole gap has been taken to infinity.  In either case even though monopoles do not explicitly appear in the effective action above their existence is manifested in the low energy theory through the quantization of the electric charge.

Applying the BW theorem to the Lagrangian in Equation \ref{u1lag} above, in addition to the thermal excitations of the gapless photon modes, we must also consider thermally excited charged particles.  Denote the gap scale of these particles by $m$.  Then the entropy density will be exponentially suppressed with a factor of $e^{-m/T}$, so that the contribution to entanglement entropy falls off as $e^{-mx}$ away from the surface.  We therefore see that, while thermal photon excitations exist far into the bulk, thermal particle excitations are only significant within a boundary layer of size $m^{-1}$ away from the surface.  In the limit of the particle gap going to infinity, particle excitations can only occur right at the surface itself.  Thus, we have a picture of the entanglement spectrum as a combination of a thermal spectrum of particles living at the boundary and a local thermal spectrum of photons living throughout the bulk.  One might naively think that we should also include thermally excited magnetic monopoles in our treatment, since these too are gapped excitations of our physical system.  However, these end up not contributing to the entropy.  This is a slightly subtle point which requires the previously mentioned embedding of the theory into a UV complete relativistic description.  More details on this procedure and on the absence of a monopole contribution can be found in Appendix C.  The electric and magnetic perspectives are complementary, rather than additive, and we may proceed considering only the electric contribution.

The two pieces of the entropy identified earlier are much clearer within this framework.  In Equation \ref{break}, we broke up the entanglement entropy into $S_{bc}$ and $S_{int}$.  Within the current local thermal picture, the first term is clearly identifiable as the thermal entropy of the boundary gas of electric particles.  The internal entropy then corresponds to the fluctuations of the strings which end at these particles on the boundary, or in other words, thermal photon excitations in the bulk.  We can also justify the earlier replacement of $S_{int,bc}$ by its average.  This entropy represents the entropy of fluctuations of strings, given the restriction that they must end on a specific particle configuration on the boundary.  But this is no significant restriction, since the temperature is arbitrarily high near the boundary, causing the strings to fluctuate wildly right near the boundary.  If we followed along a string starting from a specific point on the boundary, the wild fluctuations at the start will cause the string to almost immediately forget its starting position, so that all choices of boundary conditions are essentially equivalent for determining the photon entropy.  Thus, the entanglement entropy separates cleanly into two pieces, which we now denote as:
\begin{equation}
S = S_{part} + S_{phot}
\end{equation}
to represent the particle and photon contributions to the entanglement entropy, respectively.

\section{Universal Logarithms in the Entanglement Entropy}

\subsection{Particle Entropy}

We now wish to actually evaluate these two contributions to the entanglement entropy and see if they have any universal features which characterize the phase.  We shall first treat the particle contribution to the entanglement entropy, since it will be seen that its universal portion is the natural generalization of topological entanglement entropy (which usually characterizes gapped phases) to the present gapless case.  The universal photon contribution, on the other hand, will be essentially local in character.

To do this, we shall consider a case where the particle of minimum electric charge exists as an independent excitation.  This accounts for many $U(1)$ phases, but would seem at first to not describe the physics of phases such as the $\theta = \pi$ phase\cite{senthil}, which has dyons as the fundamental charge units.  However, as discussed in Appendix D, there is little essential difference in this case.  The universal contributions to the entanglement entropy will be exactly the same (though the precise structure of the entanglement spectrum will likely be different).  Thus, for now, we shall speak in terms appropriate to the case of an independent minimum electric charge.

We now wish to evaluate the thermal entropy of a gas of such electric charges.  While these particles in general interact through their corresponding electromagnetic fields, we are greatly aided in our quest by the fact that we are considering a deconfined phase, where the excitations of the theory are independent particles interacting through a Coulomb interaction.  Based on our local thermal perspective, these charged particles are only excited in a thin two-dimensional boundary layer near the partition.  Nevertheless, these thermally excited particles interact through the full three-dimensional Coulomb interaction ($V\sim \frac{1}{r}$) of the theory.  (Note that, while particles are excited only near the boundary, the corresponding electric field lines ending on those particles can extend into the bulk, allowing the particles to maintain their three-dimensional interactions.  Furthermore, the particles are gapped, preventing them from qualitatively modifying the long-range behavior of the gauge field.)  Now we take advantage of the fact that a thermal gas with $\frac{1}{r}$ interactions is always screened (entropic considerations causes large particle-antiparticle pairs to have favorable free energy).  Thus, when the ground state of our theory corresponds to a deconfined phase, the corresponding statement in the thermal boundary gas is that it is in a screened phase.  (A confining phase would correspond to a dipolar phase in the boundary gas, where we have tightly bound particle-antiparticle pairs.)  The particle interactions are screened, allowing us to essentially consider only a short-ranged interaction, instead of the original long-ranged Coulomb interaction.  We work deep in the screened phase, so that we can consider a contact interaction between particles on the same site as the dominant interaction.

Despite the simplicity of the model, the physics is not totally trivial.  We are still enforcing the closed loop constraint in the wavefunction.  Since no loop can end in the interior of the region, every unit of flux that goes into the region must come out at some other point.  This corresponds to the fact that the boundary gas must be neutral, having an equal number of positive and negative charges.  Thus, the partition function will be restricted to neutral configurations.  Each (unrestricted) configuration can be labeled by a set of integers $n_i$ representing the charge on boundary site $i$.  The restricted configurations in the ground state must then satisfy $\sum n_i = 0$.  Furthermore, since we are only assigning an energy cost to particles occupying the same site (a contact interaction), there is no correlation between occupation numbers on different sites, so the probability distribution for the integers $n_i$ factorizes into independent probability distributions for the occupation of each site, $f(\{n_i\}) = \prod_i f_i(n_i)$, only subject to the overall neutrality constraint.  

A simple argument shows the effect of the neutrality constraint. For simplicity let us first assume that only $n_i = 0, \pm 1$ are allowed for any single $i$, with equal probability.  Without the neutrality constraint the total number of allowed configurations of the ${n_i}$ for a lattice of $N$ sites is $3^{N}$. From random walk arguments we expect that the effect of the neutrality constraint is to reduce this by a factor $\frac{1}{\sqrt{N}}$. The entropy of this boundary gas will thus have a subleading correction $-\frac{1}{2} \ln(N)$ to the leading term proportional to $N$. Below we will provide a more detailed derivation of this result, for generic distribution $f(n_i)$.

\begin{figure*}
\centering
 \includegraphics[scale=0.6]{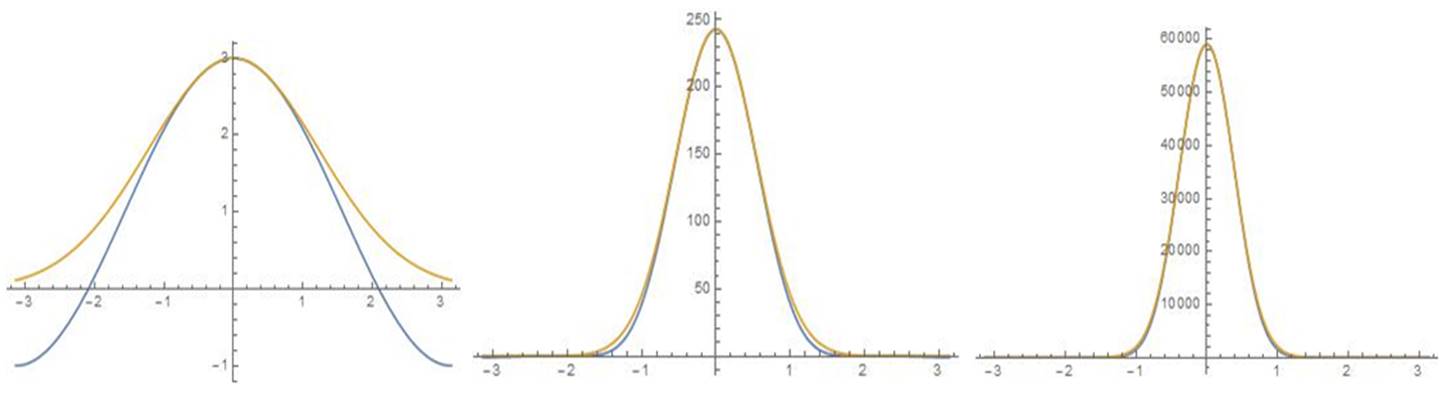}
 \caption{The first figure displays the function $\tilde{f}(b) = 1+2\cos b$ appropriate to the hard-core example in blue, along with its Gaussian approximation in yellow.  The second figure shows both functions to the fifth power, and the last figure shows them to the tenth, at which point the functions are essentially indistinguishable, with the error becoming negligible as a fraction of the maximum.}
 \label{fig:gauss}
 \end{figure*}

Taking the neutrality constraint into account, the boundary gas has a partition function as follows:
\begin{equation}
Z = \sum_{\{n_i\}}\delta_{\sum n_i} \prod_i f_i(n_i)
\end{equation}
where the delta function imposes neutrality.  We now represent the delta function via its Fourier transform, $\delta_n = \frac{1}{2\pi}\int_0^{2\pi} db e^{ibn}$, which gives us:
\begin{align}
\begin{split}
Z = \int_0^{2\pi}\frac{db}{2\pi}& \sum_{\{n_i\}} e^{ib\sum_i n_i} \prod_i f_i(n_i) = \\ \int_0^{2\pi}\frac{db}{2\pi}\prod_i \sum_{n_i}&e^{ibn_i} f_i(n_i) =
\int_0^{2\pi}\frac{db}{2\pi}\prod_i \tilde{f}_i(b) = \\ &\int_0^{2\pi}\frac{db}{2\pi}(\tilde{f}(b))^N
\end{split}
\end{align}
where $\tilde{f}_i$ is the Fourier transform of $f_i$, and in the last step we have assumed that all sites are equivalent, so that $\tilde{f}_i = \tilde{f}$ for all $i$.  $N$ denotes the total number of boundary sites, which will scale as the area of the boundary.  As a concrete example, suppose we imposed a hard-core constraint on our loops, such that only values $\pm 1$ and $0$ are allowed on each site, but each of these is equally likely.  Such a wavefunction could serve as a useful trial wavefunction for a spin-$1$ system.  In this case, we would have:
\begin{align}
\begin{split}
&Z_{h.c.}\propto \int_0^{2\pi} \frac{db}{2\pi}\prod_i \sum_{n_i = -1}^1 e^{ibn_i} = \\ \int_0^{2\pi}\frac{db}{2\pi}(1&+2\cos b)^N = {}_2F_1\bigg(-\frac{N-1}{2},-\frac{N}{2},1;4\bigg)
\label{geo}
\end{split}
\end{align}
where the integral can miraculously be done exactly in terms of a not-so-useful hypergeometric function.

For a generic distribution function $f$, there would seem to be little hope of doing the integral exactly.  However, things become very simple in the large $N$ limit.  When $N$ is large, the power of $N$ in the integrand means that the value of the integral is dominated by the values of $\tilde{f}(b)$ in the vicinity of its maxima.  First, assume that $\tilde{f}(b)$ has a unique maximum, as is the case in the above hard-core example.  This, with minor modifications, is actually a fairly generic situation.  The few pathological cases for which the following procedure will not work are discussed in Appendix F.  For the present situation, we Taylor expand $\tilde{f}(b)$ around its maximum at $b_0$ as $\tilde{f}(b) \approx c(1-\alpha (b-b_0)^2)$ for some constants $c$ and $\alpha$.  Since the integral is dominated by this behavior near the maximum, we can well-approximate the integral by replacing $\tilde{f}(b)$ by a function with an equivalent Taylor expansion near its maximum.  A convenient choice is simply the Gaussian $ce^{-\alpha(b-b_0)^2}$, which is illustrated in Figure \ref{fig:gauss}.  Using this replacement function, the partition function (for generic factorized distribution function $f$) becomes:
\begin{equation}
Z \approx \int_0^{2\pi}\frac{db}{2\pi}c^Ne^{-\alpha N(b-b_0)^2} \approx \frac{c^N}{2\sqrt{\pi\alpha N}}
\label{partition}
\end{equation}
where in the last step we let $b$ run over all real values, since the integrand is negligible away from $b=b_0$ (we can always choose the range of integration such that $b_0$ is away from the edge of the range).
It should be noted that charge quantization is actually not crucial to this argument.  If $f(n_i)$ were not restricted to integers, but could run over all real values, we would simply change to a Fourier transform on the real line, with the variable $b$ running over all real values.  The structure of the final answer will be exactly the same.  We now recall that, in terms of the partition function $Z$, the entropy is given as:
\begin{equation}
S = \beta(E-F) = -\beta\partial_\beta \log Z + \log Z
\end{equation}
Note that, since our original distribution function $f$ should depend on the inverse temperature $\beta$ (which will be determined by the lattice cutoff at the boundary and the masses of the particles), the numbers $c$ and $\alpha$ in Equation \ref{partition} will depend on $\beta$.  Restoring this dependence, the entropy is equal to:
\begin{align}
\begin{split}
S = (1-\beta\partial_\beta)(N\log c_\beta &- \frac{1}{2}\log N - \log(2\sqrt{\pi\alpha_\beta})) = \\
(\log c_\beta - \beta\frac{c'_\beta}{c_\beta})N &- \frac{1}{2}\log N + \mathcal{O}(1)
\end{split}
\end{align}
In a system with $d$ spatial dimensions, the boundary size $N$ is given by $g(\frac{L}{a})^{d-1}$, where $g$ is some number of order unity, $L$ is a characteristic linear size of the partitioning surface, and $a$ is the lattice scale.  In terms of the more commonly used variable $L$, the entropy is given (up to $\mathcal{O}(1)$ terms) as:
\begin{equation}
S = G\bigg(\frac{L}{a}\bigg)^{d-1} - \bigg(\frac{d-1}{2}\bigg)\log L
\end{equation}
where $G$ is a relatively unimportant constant (independent of $L$).  The first term represents the usual area law term of entanglement entropy, which has a non-universal coefficient $G$ due to the presence of the short-distance cutoff $a$.

The second term, however, is universal, at least at the level of independence of lattice scale.  We shall see later that it is indeed a topologically robust characterization of the phase.  This subleading logarithm originates from the factor of $\frac{1}{\sqrt{N}}$ in the partition function.  Importantly, this term in the entropy is negative, representing the reduction in entropy due to the closed loop constraint.  Seeing a loop enter the subsystem immediately gives us the extra piece of information that there is a corresponding outgoing flux at another point.  This extra information leads to a reduction in entropy.  Essentially, this contribution, and topological entanglement entropy more generally, comes from the reduced size of the Hilbert space, which reduces the number of allowable boundary conditions between regions.  As an example, in the toric code, which is characteristic of a deconfined $Z_2$ gauge theory, the ground state is an equal weight superposition of all possible loops (which are undirected in the $Z_2$ case).  In this model, the boundary values are binary, either 0 or 1.  All boundary conditions are weighted equally, as long as they are consistent with the closed loop constraint.  This constraint uniquely determines the last boundary value once we have arbitrarily picked the first $N-1$.  This leads to $2^{N-1}$ possible boundary conditions, all weighted equally.  The entanglement entropy is then given by:
\begin{equation}
\log(2^{N-1}) = (\log 2)N - \log 2
\end{equation}
The closed loop constraint restricts us to only half of the total possible boundary conditions, leading directly to the $-\log 2$ in the entanglement entropy.  Similarly, the factor of $\frac{1}{\sqrt{N}}$ in our partition function represents a reduction in the effective size of our Hilbert space.  (As an example, one can perform a large $N$ expansion of the hypergeometric function in Equation \ref{geo}, with the end result that the neutral hard-core Hilbert space is smaller than the total hard-core Hilbert space by a factor of $\frac{1}{\sqrt{N}}$.)  It is precisely this Hilbert space reduction factor of $\frac{1}{\sqrt{N}}$ which leads to the subleading logarithm.  In this respect, this subleading logarithm seems to be the generalization of topological entanglement entropy to the gapless $U(1)$ spin liquid.  Later, we shall make a more precise statement regarding this analogy with topological entanglement entropy.

For completeness, it should also be noted that the same reduction of boundary conditions is true in the confined phase.  However, when the bulk represents a confined phase, the boundary gas will be in a dipolar phase, where each particle must be very close to its antiparticle pair.  In essence, this prevents the ground state wavefunction from fully sampling the space of allowable boundary conditions.  The ground state is restricted to one particular corner of the Hilbert space and will not be a good measure of the size of the Hilbert space, so the effects of the Hilbert space restriction are not apparent.  It is only in the deconfined phase that the Hilbert space is well-sampled by the wavefunction.  Just as the $-\log 2$ is absent in the confined phase of a $Z_2$ gauge theory, the universal logarithm will not be present in the confined phase of the $U(1)$ gauge theory.  The universal logarithm shows up in the deconfined phase since each particle interacts with an average field due to all other particles, and neutrality can be enforced simply as a global constraint.  In the confined phase, the partition function would need to be written in terms of strongly interacting particle-antiparticle pairs, instead of in terms of independent particles, so the neutrality constraint would already be built into the fundamental degrees of freedom.

For later purposes, we note that, if the boundary consisted of multiple disconnected components, then we would have separate boundary gases on each surface, with separate neutrality constraints.  The entanglement entropy would be the sum of the contributions from the various connected components.  Assuming each connected component to be characterized by a similar length scale $L$, this will simply give a factor of the zeroth Betti number $b_0$, which counts the number of connected components of the partitioning surface, in front of the universal logarithm.  Thus, more generally, the particle contribution to the entanglement entropy is given by:
\begin{equation}
S_{part} = G\bigg(\frac{L}{a}\bigg)^{d-1} - b_0\bigg(\frac{d-1}{2}\bigg)\log L
\end{equation}
for some constant $G$.

\subsection{Photon Entropy}

We also need to evaluate the entropy coming from the gapless photon excitations of our system.  Intuitively, this is a very simple problem.  In the local thermal picture, the photon contribution to the entropy comes from the gapless tranverse fluctuations of strings in the bulk.  The transverse fluctuations of these gauge strings give us $d-1$ degrees of freedom at every point (where $d$ is the spatial dimension, not spacetime), so the problem should essentially be equivalent to local thermal fluctuations of $d-1$ scalar fields.  Indeed, this equivalence has been noted before \cite{rt,kabat}.  One should start with the low-energy effective action, describing standard noncompact electrodynamics in $d+1$ dimensions, without matter fields:
\begin{equation}
S = \frac{1}{2e^2}\int d^{d+1}x F^{\mu\nu}F_{\mu\nu}
\label{low}
\end{equation}
(ignoring the $\theta$ term, which is treated in Appendix D).  Earlier treatments of this problem have found that the entanglement structure of this theory is indeed equivalent to that of $d-1$ real scalar fields, up to boundary terms.  In many earlier works, the boundary terms have proven somewhat pesky in attempting to extract entanglement entropy.  Within the current framework, however, the boundary terms have a natural interpretation in terms of the particle entropy considered in the previous section, while the remaining photon contribution will be described by a theory of $d-1$ free massless scalar fields.

It should be noted that previous treatments of this problem have had conflicting results regarding the particle entropy found in the previous section.  Some authors choose to disregard the boundary term entirely \cite{rt}.  In other cases, focusing on a strictly planar entangling surface may be the limiting factor \cite{kabat}.  Since the particle contribution is proportional to the number of connected components, a concept only well-defined on a closed manifold, it is unclear if any such universal particle contribution can be identified on the infinite plane without carefully specifying the topology at infinity.  Another possibility is that the low energy action of Equation \ref{low} cannot capture the entanglement structure of the theory without a more careful treatment of the underlying Hilbert space.  These are all interesting questions which need to be sorted out at the field theory level.  But since we already have an independent way of extracting the particle contribution to the entanglement entropy, we leave these questions as an open challenge for the field theorists, and we proceed to extract the photon entropy from the bulk scalar fields.

\begin{figure*}
 \begin{minipage}[b]{0.45\linewidth}
 \centering
 \includegraphics[scale=0.5]{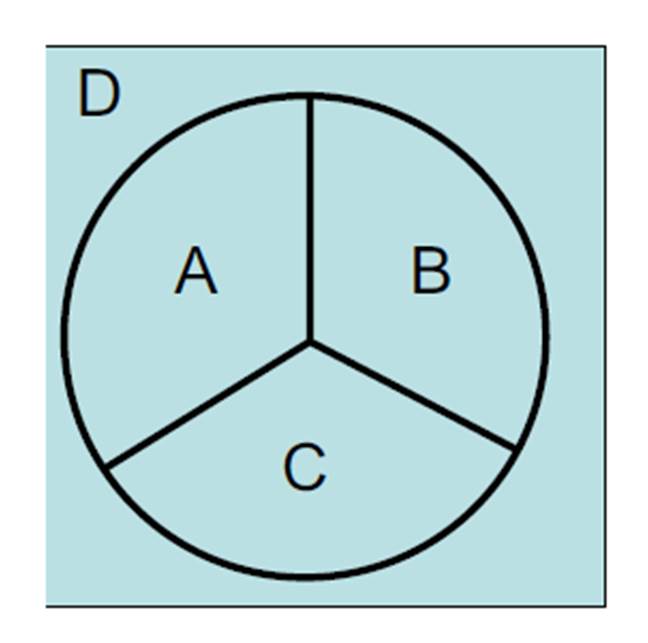}
 \caption{The Kitaev-Preskill construction allows one to isolate topological entanglement entropy in two dimensions \cite{kitaev}.}
 \label{fig:kitaev}
 \end{minipage}
 \hspace{1cm}
 \begin{minipage}[b]{0.45\linewidth}
 \centering
 \includegraphics[scale=0.5]{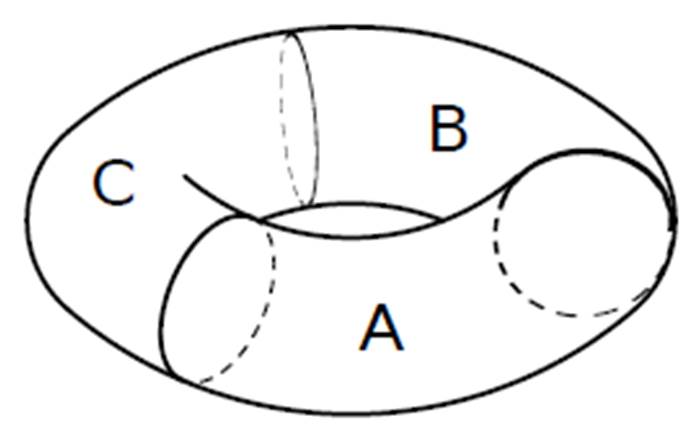}
 \caption{The Grover-Turner-Vishwanath construction generalizes the Kitaev-Preskill scheme to three dimensions and isolates the analogue of topological entanglement entropy in the $U(1)$ system \cite{grover}.}
 \label{fig:grover}
 \end{minipage}
 \end{figure*}

These scalar fields, being massless, are described by a conformal field theory.  Whereas in 1+1 dimensions a conformal field theory is characterized by a single central charge $c$, the behavior in higher dimensions is more complicated.  For concreteness, we now focus on the case of 3+1 dimensions, where a CFT is characterized by two independent central charges, denoted $a$ and $c$ (see \cite{rt} for details).  We will here take the normalization scheme such that $a$ and $c$ are equal to 1 for a real scalar field, as in \cite{mezei}.  The problem of entanglement entropy in 3+1 dimensional CFTs has been studied before, and the answer is known to be an area law with a universal subleading logarithm, $S = \alpha(L/a)^2 - \gamma \log L$ \cite{solo,rt,mezei}.  The (positive) coefficient $\gamma$ of the subleading logarithm is computed via the replica trick through differential geometric means, yielding an answer of \cite{mezei}:
\begin{equation}
\gamma = \frac{a}{180}\int_\Sigma \sqrt{h}E + \frac{c}{240\pi}\int_\Sigma \sqrt{h}I
\end{equation}
where $h$ is the induced metric on the partitioning surface $\Sigma$, and $E$ (the Euler density) and $I$ are related to the extrinsic curvature $K_{ab}$ by $E = \frac{1}{2\pi}K$ and $I = K_{ab}K^{ab}-\frac{1}{2}K^2$.  The details of the differential geometry are not overly important.  The most important fact about this answer is that both terms are given by an integral over the partitioning surface, a fact which we will return to soon.  For the case at hand of $d-1$ scalars in $d=3$ dimensions, we have $a = c = 2$, so:
\begin{equation}
\gamma_{U(1)} = \frac{1}{90}\int_\Sigma \sqrt{h}E + \frac{1}{120\pi}\int_\Sigma \sqrt{h}I
\end{equation}
The first term is given by the Gauss-Bonnet theorem as $\frac{1}{90}\chi = \frac{1}{45}(1-g)$, where $\chi$ and $g$ are the Euler characteristic and genus of the surface, respectively.  The second term is not given by a topological invariant, but it does vanish when $\Sigma$ is a sphere.  For the specific case of a sphere, we have $\gamma_{\textrm{sphere}} = \frac{1}{45}$.  Other surfaces deformed from the sphere will have corrections due to the $c$ term.

\subsection{Grover-Turner-Vishwanath Construction}

If we add together our two contributions, from particles and photons, we find the final answer for the entanglement entropy associated with partitioning surface $\Sigma$ in $d=3$ spatial dimensions (with $b_0$ connected components) to be:
\begin{equation}
S_{d=3} = \alpha\bigg(\frac{L}{a}\bigg)^2 - \bigg(b_0 + \frac{1}{90}\chi + \frac{1}{120\pi}\int_\Sigma \sqrt{h}I\bigg)\log L
\end{equation}
(up to $\mathcal{O}(1)$ terms) for some nonuniversal constant $\alpha$.  For the special case of a sphere, this answer reduces to:
\begin{equation}
S_{\textrm{sphere}} = \alpha\bigg(\frac{L}{a}\bigg)^2 - (1+\frac{1}{45})\log L
\end{equation}
We therefore see that there are two universal contributions in the subleading logarithmic behavior of the entanglement entropy, one coming from the closed loop constraint and one from the presence of a gapless photon mode.  First of all, the magnitude of the photon contribution is much smaller than the particle contribution, so the particle contribution will set the natural scale for the subleading logarithm in the entanglement entropy of a specific region.  However, it is important to note that these contributions can be separated out, regardless of magnitude, since the photon contribution is given by an integral over the surface.  Methods have been constructed precisely to eliminate such terms in favor of the topological contributions, such as $b_0$, which have no expression as an integral over the surface.  In two dimensions, Kitaev and Preskill developed the construction depicted in Figure \ref{fig:kitaev} \cite{kitaev}.  If one examines the quantity:
\begin{equation}
S_{top} = S_A + S_B + S_C - S_{AB} - S_{BC} - S_{AC} + S_{ABC}
\label{top}
\end{equation}
then one finds that all contributions given by integrals over the surface will cancel out.  However, if we had a contribution proportional to $b_0$ in the entanglement entropy, it would be present in each term of $S_{top}$ equally, since each region is connected.  With four positive terms and three negative, the end result for $S_{top}$ will isolate the $b_0$ contribution and eliminate any surface terms.

The Kitaev-Preskill construction was generalized to three dimensions by Grover, Turner, and Vishwanath, who developed the construction depicted in Figure \ref{fig:grover} (there are other possible geometries) \cite{grover}.  Using this geometry, one can define the same quantity $S_{top}$ as in Equation \ref{top}.  The photon contribution, being equal to an integral over the partitioning surface, will cancel out of this expression.  The particle contribution, however, is equal to $-\log L$ in every term of $S_{top}$.  (This assumes the geometry is characterized by a single characteristic size $L$.  We should let both radii of the torus in Figure \ref{fig:grover} be of order $L$.)  Thus, the end result for the topological portion of the entanglement entropy in a three-dimensional $U(1)$ spin liquid is:
\begin{equation}
S_{top,3} = -\log L
\end{equation}

\subsection{A Conjecture on Higher Spatial Dimensions}

One final conjecture seems worthy of mention.  In higher spatial dimensions, the particle contribution to the entanglement entropy is still an area law with a subleading logarithm, and the coefficient of the logarithm is generalized to $\frac{d-1}{2}$.  However, the low-energy CFT contribution can be more drastically altered.  In $d>3$ spatial dimensions, the entanglement entropy may behave as $\alpha L^{d-1} + \gamma L^{d-3} + ...$, so any logarithmic behavior would be dwarfed by bigger subleading corrections to the area law.  There is no sense in which the topological logarithm will represent the dominant subleading behavior of any one region.  Nevertheless, the topological logarithm should still be isolatable via some further higher-dimensional generalization of the Kitaev-Preskill scheme which eliminates boundary terms.  The scalar field theory describing the transverse fluctuations would seem to be topologically trivial, so its contribution to the entanglement entropy should be given in terms of integrals over the boundary, just as in three dimensions.  These contributions should then be eliminated by the generalized Kitaev-Preskill scheme, leaving only the topological logarithm coming from the particle contribution.  In general spatial dimension $d>3$, we should then have:
\begin{equation}
S_{top} = -\bigg(\frac{d-1}{2}\bigg)\log L
\end{equation}

\subsection{Comparison with Previous Results}

In this paper, we have found that the entanglement entropy of the $U(1)$ spin liquid (equivalently a compact $U(1)$ gauge theory) has two universal subleading logarithmic contributions.  One coefficient, $-1$, is topological, while the other coefficient, $-1/45$, arises from local physics, combining for a total of $-46/45$.  We should take a moment to compare this result with earlier literature.  The most direct calculation of the entanglement entropy is found in Reference \onlinecite{radi}, which computes that the entanglement entropy decomposes into a sum of two terms, just as we found here.  One term corresponds to the entanglement entropy of $(d-1)$ massless scalar fields, which corresponds precisely to the photon contribution identified here, and will give a contribution of $-1/45$ in three dimensions (though this is not mentioned explicitly in that work).  The other term arises from the gauge constraint on the photon (``lack of a zero mode" in that author's language), and is claimed to give a contribution of $\frac{d-1}{2}$ to the logarithm coefficient, corresponding to the topological component identified here.  The author also claims this term to be the generalization of topological entanglement entropy, just as we have claimed here.  The only discrepancy appears to be in the sign of this topological term, which seems to be positive in Reference \onlinecite{radi}, whereas we found a negative contribution here.  We note that this contribution must be negative on physical grounds, since the gauge constraint gives us an extra piece of information about the system and therefore decreases the entropy.  It therefore seems likely that the topological contribution identified in Reference \onlinecite{radi} has lost a sign somewhere.

A different line of argument is given in Reference \onlinecite{don2}.  Since the Maxwell Lagrangian is ostensibly scale invariant, one is tempted to use standard results of conformal field theory, which show that the coefficient of the logarithm is proportional to the central charge $a$, which for the case of Maxwell theory gives $-31/45$.  Reference \onlinecite{don2} goes on to find that, within their calculational framework, there is a boundary contribution of $-1/3$.  They interpret this $-1/3$ as being part of the $-31/45$, with the difference of $-16/45$ being made up by the thermal calculation of Reference \onlinecite{dowker}.  However, we believe that the correct interpretation is that such a boundary term is actually supplemental to the central charge result, giving a total of $-\frac{31}{45}-\frac{1}{3} = -46/45$, just as we found here.  The $-16/45$ result of Reference \onlinecite{dowker} is also insufficient, since it only accounts for thermally excited photons, but not for the boundary particles.

One may logically ask why the final answer for the logarithmic coefficient is not simply the result based on the central charge $a$, which would give $-31/45$.  Upon examining the standard derivations for such results (such as in Section 4.2 of Reference \onlinecite{rt}), the answer is apparent.  The entanglement entropy given by these conformal field theory derivations always yields a result which is given by an integral over the partitioning surface.  Such a result is therefore topologically trivial and will be eliminated by the Grover-Turner-Vishwanath procedure.  The standard CFT analysis is incapable of producing the topological contribution.  This is essentially because these derivations focus on the local changes in expectation values caused by local changes in the curvature, but they do not pay attention to the sensitivity of the theory to the global topology of the manifold.  For most CFTs, this is not an issue, but Maxwell theory has different topological sectors on topologically nontrivial manifolds, so the partition function is highly sensitive to topology.\cite{hfb04}  It seems likely that the standard CFT results could be modified to account for such topological effects, but the current results for CFTs can really only be trusted for topologically trivial theories.

\section{Discussion}

In this work, we have identified two different universal contributions to the subleading logarithmic behavior of the entanglement entropy of a $U(1)$ spin liquid.  One contribution, coming from the gapless photon, is essentially local in nature, given by an integral over the partitioning surface.  The other contribution, which is associated with particle excitations, serves as the natural generalization of topological entanglement entropy to the gapless $U(1)$ spin liquid.  Furthermore, this quantity seems to be the more robust characterization of the deconfined phase.  Through some fine tuning, one could possibly reach a $U(1)$ phase described by a theory with deconfined gapped charges and gapless photon, but with an alternate photon structure, such as a quadratic dispersion.  In such a theory, the universal photon contribution would not continue to exist unaltered.  The particle contribution, however, would still have the universal $-\log L$ behavior, since its presence is caused simply by the combination of deconfinement and the closed loop constraint.  Thus, this topological logarithm is a robust characterization of a deconfined $U(1)$ phase.  We note that similar results have recently been arrived at through a different method than the one applied here \cite{radi}.

We focused on the entanglement entropy of the $(3+1)$-dimensional $U(1)$ quantum spin liquid in this paper. In the absence of some global symmetry in the   the underlying microscopic Hamiltonian there is a unique such phase. 
The presence of global symmetry will subdivide this into many symmetry-enriched spin liquid phases \cite{senthil}. The universal terms we found in the entanglement entropy will be the same for all these symmetry enriched phases. However we expect that the entanglement spectrum will be able to distinguish them from each other. For the future it will thus be interesting to study the entanglement spectrum of these spin liquid phases. 

It would also be interesting to examine the entanglement entropy of other gapless phases to see if universal ``topological" contributions similar to the ones we found arise.  Apart from the example discussed here, where the universal piece is logarithmic, we also have the examples of discrete gauge theories coupled to gapless matter fields, where there is a universal constant contribution.  Perhaps such contributions to the entanglement entropy  could be a useful tool for  characterizing  long-range entanglement in gapless phases.  Also, the particle logarithm identified in the present paper is a direct consequence of the existence of deconfined particles.  It is interesting to consider if the result could be phrased in terms of an effective ``quantum dimension" of the deconfined particles, just as the topological entanglement entropy in topologically ordered phases is given in terms of the quantum dimension of particles.  However, whether or not such a concept can be made mathematically precise remains to be seen.

\emph{Note added in proof}.  In this paper, we have shown that the entanglement entropy of a U(1) spin liquid separates into two contributions, both leading to universal logarithmic terms in the entanglement entropy. The most notable piece is a topological contribution, coming from the particle degrees of freedom, and can be understood as arising from a thermal distribution of particles living on the entanglement cut. This logarithm takes the form −log(L). The other is a local contribution, coming from the photon degrees of freedom, and can be understood as a local thermal ensemble of photons residing in the bulk. In the present paper, we claimed that such a thermal distribution of photons is equivalent to that of two real scalar fields, giving a contribution of −(1/45) log(L) on a sphere. However, several recent works [43,44] have drawn our attention to a more exact treatment of the thermal photon problem [30], where it is shown that the correct form of the bulk term is actually −(16/45) log(L) on the sphere, which differs from the two scalar result for subtle reasons. We therefore wish to note that the correct local term seems to be given by −(16/45) log(L). However, this does not modify the topological contribution identified in this paper, which remains −log(L).

\section*{Acknowledgments}

The authors would like to thank Sagar Vijay and Liujun Zou for insightful discussions on the present work, and also Tarun Grover for early encouragement.  This work was supported by NSF DMR-1305741.  This work was also partially supported by a Simons Investigator award from the
Simons Foundation to Senthil Todadri.

\section*{Appendix A: Eliminating $a_0$}

In the standard Lagrangian formulation of the low-energy theory of a deconfined $U(1)$ phase, the variables used to describe the theory are the components of a (noncompact) spacetime vector $a_\mu = (a_0,\vec{a})$. However, the action for the theory only depends on the field strength tensor $F_{\mu\nu} = \partial_\mu a_\nu - \partial_\nu a_\mu$.  The first thing to take note of is the fact that $F_{\mu\nu}$ contains no $\partial_0 a_0$ term, so the variable $a_0$ has no dynamics whatsoever.  This would seem to hint that $a_0$ is not really a dynamical degree of freedom at all.

Another hint for this comes from lattice gauge theory defined on a spacetime lattice, where time is discrete.  In this case, the spatial vector potential $\vec{a}$ is defined on the spatial links of the lattice, but the timelike component $a_0$ is actually defined on the timelike links of the lattice.  This would seem to be a very unnatural place for a degree of freedom to live.  In a quantum mechanical system, we specify the state of the system at a specific moment of time.  The variable $a_0$, however, does not exist on a specific time slice of our system, but rather on a set of links connecting two time slices.  This would also seem to indicate that $a_0$ is not really a degree of freedom in our system, but rather serves as a constraint on the time dynamics of the system.

In fact, $a_0$ can be integrated out of the low energy field theory to yield the Gauss's law constraint, as follows.  For the Lagrangian $\mathcal{L} = \frac{1}{2}F_{\mu\nu} F^{\mu\nu}$ (with the normalization chosen for later convenience), we write the partition function for our system as:
\begin{equation}
Z = \int \mathcal{D}a_0\mathcal{D}\vec{a} \exp\bigg[ \frac{1}{2} i\int dx ((\partial_0 \vec{a}-\nabla a_0)^2 - (\nabla \times \vec{a})^2)  \bigg]
\end{equation}
We now introduce an integration over an auxiliary field $\vec{E}$ to make the exponent linear in $a_0$, like so:
\begin{align}
\begin{split}
Z = \int \mathcal{D}&a_0\mathcal{D}\vec{a}\mathcal{D}\vec{E}\exp\bigg[ \\ &\frac{1}{2} i\int dx (2\vec{E}\cdot(\partial_0 \vec{a}-\nabla a_0) - (E^2 + B^2))\bigg]
\end{split}
\end{align}
where we have defined $B=\nabla \times \vec{a}$.  We can now integrate the $\nabla a_0$ term by parts and integrate out $a_0$ to obtain:
\begin{align}
\begin{split}
Z = \int &\mathcal{D}a_0\mathcal{D}\vec{a}\mathcal{D}\vec{E} \exp\bigg[ \\
&\frac{1}{2} i\int dx (2\vec{E}\cdot\partial_0 \vec{a}- 2(\nabla\cdot\vec{E}) a_0 - (E^2 + B^2))\bigg] = \\
\int\mathcal{D}\vec{a}&\mathcal{D}\vec{E} \delta_{\nabla\cdot \vec{E}}\exp\bigg[i\int dx (\vec{E}\cdot\partial_0 \vec{a} -\frac{1}{2} (E^2 + B^2))\bigg]
\end{split}
\end{align}
This path integral now represents the Hamiltonian formulation of a theory with canonical conjugate variables $\vec{a}$ and $\vec{E}$, and with Hamiltonian given by $\frac{1}{2}(E^2+B^2)$, plus an infinite energy penalty for a nonzero value of $\nabla\cdot \vec{E}$.  (Many would prefer to regard $\nabla\cdot\vec{E}\neq 0$ states as simply nonexistent in the pure gauge theory.  However, as discussed in the main text, it is more convenient to regard these states as particle states, which are simply gapped out to high energies.)  Thus, we see that we can equivalently formulate the problem purely in terms of a spatial vector potential $\vec{a}$ and its conjugate momentum $\vec{E}$.  All of the physically meaningful entanglement properties will be captured by the Hilbert space of the spatial gauge degrees of freedom on the spatial links.

\section*{Appendix B: Projecting Direct Product States}

We now demonstrate the assertion, claimed in the text, that a direct product state, when projected onto the zero particle sector, will have an entanglement spectrum in one-to-one correspondence with the electric boundary conditions.  In order to perform such a projection, we must remove both electric and magnetic monopole configurations from the wavefunction.  It is convenient to perform the magnetic projection first.  In terms of magnetic flux, a magnetic monopole represents a point near which the flux is large, in the sense of large deviations from its minimum energy values of $2\pi n$, for integer $n$.  In order to project away magnetic monopoles, we must project away large values of magnetic flux.  One simple way to insure this is to simply project away large values of the vector potential itself, keeping only configurations in which the vector potential $a$ has small fluctuations around $a=0$.  This may seem to be over-projecting, since there are many gauge-equivalent configurations of $a$ which have large $a$ but still have small magnetic flux.  However, as we shall see shortly, the projection onto zero electric charge naturally restores equal weight to all gauge-equivalent configurations, so we are free to pick a gauge with small $a$ at this stage.  Thus, in order to project away magnetic monopoles, it is sufficient to project linkwise onto the subspace with small $a$.  This can be defined with some arbitrary cutoff $a_{max} \ll \pi$.

Thus, the magnetic projection can be performed through a product of linkwise projections, $P_{mag} = \prod_i P_{a,i}$.  Now we write the wavefunction for our system in direct product form as:
\begin{equation}
|\Psi\rangle = |\psi\rangle_A |\phi\rangle_B
\end{equation}
for some partition of the links of the system into subsystems $A$ and $B$.  Importantly, all links are either in $A$ or in $B$ unambiguously.  Thus, the magnetic projection operator factorizes as $P_{mag} = \prod_{i\in A}P_{a,i}\prod_{i\in B} P_{a,i} \equiv P_A P_B$.  Acting with this projector on our direct product state gives:
\begin{equation}
P_{mag}|\Psi\rangle = (P_A|\psi\rangle_A)(P_B|\phi\rangle_B) \equiv |\psi'\rangle_A |\phi'\rangle_B
\label{magproj}
\end{equation}
which is still a direct product state between the projected states $|\psi'\rangle_A$ and $|\phi'\rangle_B$.

With the magnetic monopoles tamed, we must now project away electric particles.  We first demonstrate that this restores equal weight to all gauge-equivalent states.  Going over to path integral notation for convenience, a generic wavefunction can be written in the electric basis as:
\begin{equation}
|\Psi\rangle = \int \mathcal{D}E f(E)|E\rangle
\end{equation}
which can then be Fourier transformed as:
\begin{equation}
|\Psi\rangle = \int\mathcal{D}E\mathcal{D}a\, f(E) e^{i\int E\cdot a}|a\rangle
\end{equation}
We can project onto the closed loop (no electric particle) subspace by representing the corresponding delta function as a Fourier integral:
\begin{align}
\begin{split}
P|\Psi\rangle& = \int \mathcal{D}E \delta(\nabla \cdot E)f(E)|E\rangle \propto \\
\int& \mathcal{D}E\mathcal{D}b e^{i\int b\nabla \cdot E} f(E)|E\rangle = \\
\int& \mathcal{D}E\mathcal{D}b e^{-i\int E\cdot \nabla b}f(E)|E\rangle = \\
\int \mathcal{D}E&\mathcal{D}b\mathcal{D}a f(E) e^{i\int E\cdot (a -\nabla b)} |a\rangle = \\
\int\mathcal{D}E&\mathcal{D}a f(E) e^{i\int E\cdot a} \bigg(\int\mathcal{D}b|a+\nabla b\rangle\bigg)
\end{split}
\end{align}
where we have performed an integration by parts in the second line, and we shifted the $a$ variable by $\nabla b$ in the final line.  We see that the end result of the projection is the same result as the original state, but with the basis $|a\rangle$ replaced by an equal weight superposition of all gauge-equivalent states, $\int \mathcal{D}b |a+\nabla b\rangle$.  This establishes our earlier claim that electric projection restores equal weight to gauge-equivalent states, justifying our magnetic projection procedure.  Furthermore, projecting onto the zero electric charge sector does not yield any configurations with different magnetic flux from the original wavefunctions, so this projection respects the absence of magnetic monopoles.

We now wish to explicitly electrically project the state in Equation \ref{magproj}, which has already been magnetically projected.  To do this, we note that Gauss's law, $\nabla \cdot E = 0$, is defined on the sites of the lattice, meaning that the links touching each site should in total carry out as much flux as they bring in.  The electric projector can therefore be written as a superposition of site projectors, $P_{elec} = \prod_{i\in sites}P_i$.  Unlike links, not all sites can be associated with either $A$ or $B$.  There are those sites totally in $A$ and those totally in $B$, but there are also those points on the boundary, which define the partition between $A$ and $B$.  Our projector can then be written as $P_{elec} = P_AP_BP_\partial$, where $P_A$ acts only on region $A$, $P_B$ acts only on region $B$, and $P_\partial$ acts on the boundary sites (and all three factors commute).  The final projected states can then be written as:
\begin{align}
\begin{split}
P|\Psi\rangle = P_{elec}P_{mag}&|\Psi\rangle = P_\partial (P_A|\psi'\rangle_A)(P_B|\phi'\rangle_B) \\
&\equiv P_\partial (|\psi''\rangle_A |\phi''\rangle_B)
\end{split}
\end{align}
In the end, all we are left with is a boundary projection acting on a direct product state, $|\psi''\rangle_A|\phi''\rangle_B$.  To perform the last projection, we first expand the wavefunctions on each side in a basis of eigenstates of electric flux through the boundary:
\begin{equation}
|\psi''\rangle_A = \sum_{bc} c_{bc}|bc\rangle
\end{equation}
where $bc$ denotes boundary conditions, and a similar expansion holds for $|\phi''\rangle$.  Call its expansion coefficients $c'_{bc}$.  The final wavefunction becomes:
\begin{align}
\begin{split}
P|\Psi\rangle =& P_\partial \sum_{bc,bc'} c_{bc}c'_{bc'}|bc\rangle_A |bc'\rangle_B = \\
&\sum_{bc} (c_{bc}c'_{bc})|bc\rangle_A |bc\rangle_B
\end{split}
\end{align}
where the final projection has enforced equality of boundary conditions.  This final wavefunction is already explicitly in Schmidt form, with Schmidt coefficients $c_{bc}c'_{bc}$, and each Schmidt coefficient corresponds to a unique boundary condition.  We have therefore demonstrated that a direct product state, after projecting away electric and magnetic monopoles, will yield a state with entanglement spectrum in one-to-one correspondence with boundary conditions.

\section*{Appendix C: On the Nature of Magnetic Monopoles}

In the main text, we have shown that the entanglement spectrum of the $U(1)$ spin liquid is described by a photon contribution, and also a contribution which can be naturally associated with a boundary theory of electric particles.  This conclusion was reached through two separate perspectives, both at the level of the wavefunction and at the level of the local thermal picture.  Both of these frameworks might suggest that the magnetic field ought to be treated on equal footing, and that at the end of the day we ought to have a separate topological logarithm coming from magnetic neutrality.  In fact, this is not the case, as we shall now describe for both frameworks separately, first for the wavefunction picture and then for the local thermal picture.

\subsection*{Wavefunction Picture}

At the level of the wavefunction, we enforced the Gauss's law constraint by matching up electric boundary conditions between the two regions of our bipartition.  Should we not similarly match up magnetic boundary conditions in order to ensure the absence of magnetic monopoles?  The issue here comes from the fact that electric and magnetic fields do not commute.  Actually, parallel electric and magnetic fields do commute (recalling that the electric field is defined on links whereas the magnetic field is defined on plaquettes, or equivalently dual lattice links).  We could therefore simultaneously specify the normal components of both the electric and magnetic fields at the boundary.  However, this is not sufficient.  In order to ensure the absence of electric charge on a site, it is not enough to simply match up the electric field on links normal to the surface.  One must also keep track of all the flux being carried by the links running along the surface.  Similarly, matching up magnetic boundary conditions would require knowledge of both normal and transverse magnetic flux at the surface.  Due to the non-commuting nature of electric and magnetic fields, it is not possible to specify all of these quantities at the same time.  In fact, if we examine a piece of our wavefunction in a fixed flux sector, we note that the act of fixing a specified electric field $E$ will automatically lead to large fluctuations in the vector potential $a$, leading to magnetic monopole configurations.  In short, the decomposition of the wavefunction into its electric flux sectors does not respect the absence of magnetic monopoles.  Of course, the final wavefunction must still be monopole free, but this will rely on a cancellation of monopole configurations between different electric flux sectors.  We therefore see that it is not possible to label the elements of the entanglement spectrum by both electric and magnetic boundary conditions.  Furthermore, in the example of a projected direct product state, we have explicitly shown that the electric boundary condition is sufficient to label the entanglement spectrum, without mention of any magnetic boundary conditions (see Appendix B).

One may be slightly bothered by the seeming asymmetry between electric and magnetic fields, when the low-energy physics of the $U(1)$ spin liquid has an emergent electric-magnetic duality.  It would seem that the entanglement structure also has a similar dual description.  It is important to remember that we have chosen the links to represent the fundamental variables of our system.  Any partitioning of these links will automatically end up breaking up some plaquettes at the boundary, and not every plaquette will belong uniquely to a specific subsystem.  Thus, it is impossible to even describe the magnetic flux in the vicinity of the boundary as a function of information accessible on one side of the partition.  However, one could imagine a dual description, where we take our fundamental variables to be the plaquettes (or equivalently, links on the dual lattice).  If we then partitioned our system in terms of plaquettes, it would be the links which are ambiguous at the boundary.  If we took this to be the setup of our system, the natural thing would be to have an entanglement spectrum in one-to-one correspondence with magnetic boundary conditions, rather than electric boundary conditions, which would not play a significant role in the entanglement structure.  At the end of the day, we should get the same topological logarithm as in the present analysis.  Thus, we see that the electric and magnetic perspectives are dual, rather than additive, and there is only a single topological logarithm coming from the particle sector of the theory.

\subsection*{Local Thermal Picture}

It is also important to understand the absence of a magnetic contribution at the level of the local thermal interpretation of the Bisognano-Wichmann theorem.  In section 3, we noted that the low energy effective theory of our system:
\begin{equation}
S = \int F_{\mu\nu}F^{\mu\nu}
\label{ff}
\end{equation}
has an emergent relativistic symmetry.  We can therefore apply the Bisognano-Wichmann result to obtain the entanglement Hamiltonian for this system, which has a useful interpretation in terms of a local thermal picture where the temperature varies as $T\sim \frac{1}{x}$ ($x$ being the distance away from the partitioning surface).  However, to apply the local thermal perspective, we must decide what states are actually thermally excited within this picture.  The full Hilbert space of our original lattice theory contains gapless photons, gapped (deconfined) electric charges, and gapped (deconfined) magnetic monopoles.  However, the low energy action above would naively seem to indicate that the only degree of freedom is the gapless photon.  But as we have argued in the main text, in order to be consistent with emergence from a local tensor product Hilbert space (as in a spin liquid), we are forced to consider the endpoints of electric strings in our Hilbert space.  In other words, even when we have taken the mass of the electric particles to infinity, we still must consider these particles as existing within the Hilbert space.  The action \ref{ff} should really be thought of as a limiting case of a gauge field coupled to (electrically) charged matter fields, where we take the mass of the electric particles to be very large.  Since the temperature profile of the local thermal picture, $T \sim \frac{1}{x}$, grows arbitrarily large near the boundary, these particles will always be excited in some thin layer at the boundary.

We have now established that both photons and electric particles must be taken into account in the local thermal perspective.  Indeed this motivated the use of the theory in Equation \ref{u1lag} with a non-compact gauge field. But what about magnetic monopoles?  Naively, one might think that we have two separate thermal boundary gases, one of electric charges and one of magnetic charges, leading to two separate logarithms due to the neutrality constraint in each gas.  However, the logic of the previous section hints that this perspective is somehow flawed and that the magnetic monopoles should not contribute separately.  This is indeed the case.  There are multiple ways to see this.  At a straightforward (but possibly too naive) level, one could say that, while the physical theory of our system has magnetic monopoles, the relativistic low-energy effective action of Equation \ref{ff} is that of a \emph{noncompact} $U(1)$ gauge field.  In going to this low-energy theory, the electric particles are still kept in the Hilbert space in order to maintain the tensor product structure.  However, the action has apparently ``forgotten" the original compactness and done away with magnetic monopoles.  However note that we took the electric matter to have quantized charge. This quantization is a low energy manifestation of the existence of magnetic monopoles in a UV completion of the theory.

The issue with this argument is that the low-energy theory of a noncompact $U(1)$ gauge field coupled to charged matter is not a UV complete theory.  At high energies, the theory flows to strong coupling (the ``Landau pole" issue).  The high-energy behavior of this low-energy field is not really well-defined, so it is not surprising that we are running into ambiguity issues regarding monopoles when trying to consider the theory at arbitrarily high temperatures.  We are interested in theories where the UV completion is achieved through a lattice model.  This of course breaks Lorentz invariance and complicates the use of the BW theorem to arbitrarily high energies. However since we are interested in universal properties we are free to choose any other UV completion.  Luckily, there is a very simple way to get a relativistic UV complete field theory which reproduces the desired low-energy behavior.  We can do this by regarding the $U(1)$ theory as descending from some non-abelian gauge theory (say $SU(2)$) via symmetry breaking by the Higgs mechanism.  Non-abelian gauge theories exhibit asymptotic freedom and can be unambiguously described at high energies.  Such a field theory naturally incorporates the compact nature of the Hilbert space in a manifestly relativistic manner (in the form of t'Hooft-Polyakov monopoles).  We should then be able to apply the Bisognano-Wichmann result and the local thermal picture unambiguously to the symmetry-broken non-abelian gauge theory.  Since the low-energy theory is the same as in our compact $U(1)$ gauge theory, we expect the universal aspects of the entanglement entropy to be the same.

We therefore apply the local thermal picture to the symmetry broken non-abelian gauge theory.  This theory has two important energy scales: the mass $m$ of electric particles, which is an independently tunable parameter, and the mass $M$ of magnetic monopoles, which is set by the Higgs scale of the theory.  Above the Higgs scale $M$, the full non-abelian gauge symmetry is restored.  Since the high-energy details of the theory are not important for extracting universal features, we can take $m \ll M$ without loss of generality.  We therefore see that there are three relevant regions in the local thermal analysis of the non-abelian gauge theory.  In the bulk of the system, temperatures are very small, $T < m$, and only the gapless photon is thermally excited.  Moving closer to the boundary, in the region corresponding to the intermediate temperature scale $m < T < M$, we have a screened Coulomb gas of thermally excited electric particles.  Moving even closer to the boundary, we eventually hit a temperature $T>M$ where magnetic monopoles are excited.  However, at this temperature, the system ``forgets" about the Higgs symmetry breaking and we are restored to the full non-abelian gauge theory.  We will then be in a high-temperature phase of a non-abelian gauge theory, which is known to take the form of a quark-gluon plasma.  It is unclear how to extract the entropy of this quark-gluon plasma, since the non-abelian gauge structure does not provide us with a simple neutrality condition.  Luckily, there is no need for such a calculation.  This high-$T$ quark gluon plasma in the local thermal description of the entanglement Hamiltonian is universal to all phases of the non-abelian gauge theory including the confined phase. Consequently the contribution from this  region cannot distinguish the Higgs phase we are interested in from trivial phases. Thus this region will  only contribute to the area law term and will not  affect the universal particle logarithm coming from the screened boundary gas of electric charges of the low energy $U(1)$ gauge theory. 

We thus conclude, in agreement with our previous arguments, that the magnetic monopoles do not lead to an independent universal logarithm in the entanglement entropy.

\section*{Appendix D: $\theta$ Terms in the $U(1)$ Action}
 
Throughout the bulk of this work, we have taken the low-energy effective action for the three-dimensional $U(1)$ spin liquid to be standard noncompact electrodynamics:
\begin{equation}
S = \int d^4x \frac{1}{2e^2}F^{\mu\nu}F_{\mu\nu}
\end{equation}
However, in general, there is another allowable term, referred to as the $\theta$ term:
\begin{equation}
S = \int d^4x \bigg(\frac{1}{2e^2}F^{\mu\nu}F_{\mu\nu} + \frac{\theta}{32\pi^2} \epsilon^{\mu\nu\lambda\sigma}F_{\mu\nu}F_{\lambda\sigma}\bigg)
\end{equation}
The theta term will arise generically if there is no time reversal symmetry. It can arise microscopically by a change of the Hamiltonian describing the electric charge.  It has the effect that the monopoles now acquire fractional electric charge $\frac{\theta}{2\pi}$.  Note that what we call electric charge and what we call magnetic charge is arbitrary in the absence of time reversal. With any given convention, in general,  there will be a lattice of electric charge and magnetic monopole excitations. We can always choose a basis in this lattice  so that there is a pure electric charge. The monopole excitations will then appear as dyons and will live in an axis tilted at some angle to the `electric' axis. Thus any theory with a non-zero $\theta$ can be  viewed as a theory with gapped pure electric charges, gapped dyons, and a gapless photon. Now as argued in previous sections the entanglement entropy is determined by the theory obtained by integrating out the monopoles. In the absence of monopoles the $\theta$ term is a total derivative and can be ignored. Thus as far as the entanglement entropy is concerned the theory with a non-zero $\theta$ is not different from that at $\theta = 0$. 

In the presence of time reversal symmetry $\theta = 0$ or $\theta = \pi$. The latter is realized if the electric charges are Kramers fermions and form a topological insulator. Though the argument above for the entanglement entropy will continue to hold it is clear from this connection that the entanglement {\em spectrum} will distinguish $\theta = 0$ and $\theta = \pi$.

\section*{Appendix E: A Comment on Non-Separable Theories}

In the present work, we have considered a $U(1)$ spin liquid with gapless photon and gapped particles.  In this theory, we have found that entanglement entropy separates cleanly into a contribution from the gapless photon and a topological term which can be associated with the gapped particles.  The separation is particularly clear in the local thermal picture, where gapless photon excitations occur throughout the bulk, whereas thermally excited particles exist only at the boundary.

However, there exist other theories where such a clean separation between the gapless contribution and the topological contribution do not occur.  For example, the model considered in \cite{sachdev} has both a topologically nontrivial structure and gapless modes.  However, it is found that the universal contribution to the entanglement entropy is not simply the sum of these two contributions.  The distinguishing feature of models such as this, as compared to the model considered here, is that it is the particles which are gapless, whereas we have here considered only a gapless gauge mode, $i.e.$ the photon.  As we have discussed in this work, the topological part of the entanglement entropy comes from the deconfined particle excitations, which have a global neutrality constraint giving rise to topological entanglement entropy.  The gapless photon, on the other hand, has no such topologically nontrivial structure.  In models such as in \cite{sachdev}, the deconfined particles themselves are the gapless modes.  We expect these particles to contribute some sort of ``topological" contribution to the entanglement entropy, but we also expect some extra entanglement structure due to their gaplessness.  It is far from clear that the entanglement contribution of these gapless particles should be simply a sum of terms due separately to their gaplessness and due to their ``topologicalness."  In fact, the work in \cite{sachdev} indicates that such a separation does not hold.  It would be highly interesting to attempt to apply a Kitaev-Preskill sort of scheme to these models to see what comes out.  Or perhaps the Kitaev-Preskill scheme itself will need modification to correctly extract the interesting physics of such models.  There is much to explore.

\section*{Appendix F: Failures of the Fourier Integral Method}

In the main text, we have worked with a distribution function $\tilde{f}(b)$ with a unique maximum, with smooth behavior in the vicinity of that maximum.  More generally, we could easily extend this to include functions with a discrete number of smooth maxima, giving essentially the same behavior, since cross terms between maxima would be insignificant in the large $N$ limit.  The most significant exception occurs when $\tilde{f}(b)$ has a sharp delta function peak, as would occur when $f(n_i) = 1$ for all $n_i$.  This represents a state in which there is no energy penalty for particle overlap, and all boundary conditions of arbitrarily high flux are equally weighted.  In this case, the entanglement entropy diverges and there is no sensible notion of topological entanglement entropy.  This wavefunction is not totally irrelevant, since it describes the ground state of the $U(1)$ Hamiltonian with the electric field term tuned all the way to zero.  However, this seems to be a singular limit, since any regularization of $f(n_i)$ will give the subleading logarithm.

This is the most significant difference between the $U(1)$ gauge theory and the discrete case, such as a $Z_n$ gauge theory.  In a discrete gauge theory, one can carry through an analogous analysis in terms of a Fourier transform on a finite group.  The Fourier transform is discretely defined, preventing us from performing any sort of Gaussian expansion around the maximum.  In this case, when the electric field term of the Hamiltonian goes to zero, we end up with a Kronecker delta instead of a Dirac delta function in $\tilde{f}(b)$.  This Kronecker delta will then give us exactly the $-\log n$ topological entanglement entropy.  Thus, in the discrete case, we can feel free to let the coefficient of the electric field term be tuned all the way to zero, giving us essentially a Kitaev model (a generalization of the toric code).  In the $U(1)$ gauge theory, on the other hand, a correct extraction of the entanglement entropy requires regularization by retaining an energy cost for large electric fields.  Considering the fact that a $Z_n$ gauge theory has $-\log n$ topological entanglement entropy (in any dimension) and the $U(1)$ gauge theory has a $-\frac{d-1}{2}\log L$ contribution (which we have argued to be topological), it may have been tempting to view $U(1)$ gauge theory as a large $n$ limit of $Z_n$ gauge theory, with $n$ cut off at order $L^{(d-1)/2}$.  However, the use of the Gaussian expansion for the $U(1)$ theory seems distinctly different from the procedure in the $Z_n$ case, and it is questionable if any such large $n$ identification holds.

While a generic regularized function $\tilde{f}(b)$ in our $U(1)$ theory will have quadratic behavior near its maximum, it is also entertaining to consider the possibility that, through some fine-tuning, it may be possible to engineer a wavefunction such that the behavior of $\tilde{f}(b)$ is not quadratic near its maximum, but rather a higher (even) power, as $\tilde{f}(b) \approx c (1-\alpha(b-b_0)^\eta)\approx ce^{-\alpha(b-b_0)^\eta}$.  The factor of $\frac{1}{\sqrt{N}}$ in the partition function is replaced by $N^{-1/\eta}$, so the subleading logarithm becomes $-\frac{1}{\eta}\log N = -\frac{d-1}{\eta}\log L$.  The generic case of course has $\eta = 2$, but it is possible that a fine-tuned model may have a ground state with this modified subleading logarithm.  Another less interesting failure of the Fourier integral method is when $\tilde{f}(b)$ has no maxima, but rather is a constant function.  This would correspond to $f(n_i)$ being a delta function at zero charge (other charges could not satisfy the neutrality constraint), which would only occur in the wavefunction for a confined phase and is not of concern to us here.


\begin{thebibliography}{9}
\bibitem{foot}
With some caveats, to be discussed in a later section, we find that the answer in general dimension takes the form $\gamma_{top} = \frac{d-1}{2}$.
\bibitem{foot2}
If the reader is uncomfortable with this definition of matter and prefers to regard non-gauge-invariant states as unphysical, then the present work can simply be regarded as embedding the physical Hilbert space of gauge-invariant states in the larger tensor product Hilbert space of both physical gauge-invariant states and non-physical non-gauge-invariant states.  The conclusions will be the same.
\end{thebibliography}
\end{document}